\documentclass{article}

\usepackage{PRIMEarxiv}

\usepackage[utf8]{inputenc} 
\usepackage[T1]{fontenc}    
\usepackage{hyperref}       
\usepackage{url}            
\usepackage{booktabs}       
\usepackage{amsfonts}       
\usepackage{nicefrac}       
\usepackage{microtype}      
\usepackage{lipsum}
\usepackage{fancyhdr}       
\usepackage{graphicx}       
\graphicspath{{media/}}     

\RequirePackage{amsthm,amsmath,amsfonts,amssymb}
\RequirePackage{graphicx,color,xcolor,subfigure,subcaption}
\RequirePackage{cleveref,bm,booktabs,float,soul}
\usepackage[authoryear]{natbib}

\newcommand{\iid}{\stackrel{\mbox{\tiny iid}}{\sim}}
\newcommand{\ind}{\stackrel{\mbox{\tiny ind}}{\sim}}

\newcommand{\calN}{\mathcal N}

\title{Bayesian source apportionment of\\ spatio-temporal air pollution data 
}

\author{
  Michela Frigeri \\
  Politecnico di Milano \\
  Milan, Italy\\
  \texttt{michela.frigeri@polimi.it} \\
   \And
  Veronica Berrocal \\
  University of California Irvine\\
  Irvine (CA), USA\\
  \texttt{vberroca@uci.edu} \\
  \And
  Alessandra Guglielmi \\
  Politecnico di Milano \\
  Milan, Italy\\
  \texttt{alessandra.guglielmi@polimi.it} \\
}

\begin{document}
\maketitle

\begin{abstract}
Understanding the sources that contribute to fine particulate matter (PM$_{2.5}$) is of crucial importance for designing and implementing targeted air pollution mitigation strategies. Determining what factors contribute to a pollutant's concentration goes under the name of source apportionment and it is a problem long studied by atmospheric scientists and statisticians alike.
In this paper, we propose a Bayesian model for source apportionment, that advances the literature on source apportionment by allowing estimation of the number of sources and accounting for spatial and temporal dependence in the observed pollutants' concentrations. Taking as example observations of six species of fine particulate matter observed over the course of a year, we present a latent functional factor model that expresses the space-time varying observations of log concentrations of the six pollutant as a linear combination of space-time varying emissions produced by an unknown number of sources each multiplied by the corresponding source's relative contribution to the pollutant. Estimation of the number of sources is achieved by introducing source-specific shrinkage parameters.
Application of the model to simulated data showcases its ability to retrieve the true number of sources and to reliably estimate the functional latent factors, whereas application to PM$_{2.5}$ speciation data in California identifies 3 major sources for the six PM$_{2.5}$ species.
\end{abstract} 

\bigskip
\noindent
\bigskip



\section{Introduction}
\label{sec:introduction}

Epidemiological evidence on the deleterious health effects of PM$_{2.5}$ has been assembled since the 1990's when the first landmark air pollution health studies, such as the Harvard Six Cities study, were launched \citep{dockery1993,schwartz1991,schwartz1992}. Since then, the literature on the human health effects of air pollution has grown remarkably providing such a striking evidence that in 2022 the National Academies of Science, Medicine and Engineering authored a report examining the ``weight of evidence'' framework employed by the U.S. Environmental Protection Agency \citep{epaisa} to determine causality in the relationship between air pollution and adverse health effects \citep{nasem2022}.
Given these findings, it is clear that reducing the level of air pollutants' concentrations is a public health priority.

Fine particulate matter or PM$_{2.5}$ is one of the pollutants most studied by epidemiologists and environmental health scientists. It is a mixture of various components, often referred to as "\textit{species}", which arise as a consequence of natural processes, such as volcanic eruptions, dust storms, plants respiration, etc., as well as a result of human activities. The contribution of these \textit{sources} of pollution to the overall level of PM$_{2.5}$ varies spatially and temporally, leading to a composition of PM$_{2.5}$ that is heterogeneous in space and time \citep{bell2007,bell2009}.
Understanding what processes and activities produce emissions that contribute to increased concentration of PM$_{2.5}$ is of great importance when trying to develop and implement targeted interventions aimed at improving air quality.   

Source apportionment deals exactly with this problem, namely identifying and quantifying the sources that contribute to measured concentrations of one or multiple air pollutants at monitoring locations. Different research communities have tackled this problem, developing methods that vary from community to community and are often designated with an assortment of names. 
Atmospheric scientists and environmental engineers mostly use deterministic chemistry-transport models (CTM) and sensitivity analysis or emission reduction impacts (ERI) methods to determine the contribution of a particular source to the local pollution, or they use CTM's to follow the atmospheric fate of tagged sources in a technique that goes under the name of Particle Source Apportionment Technology \citep{burr2011,kwok2013,kitagawa2021}. 

Stated as a statistical problem, the source apportionment problem can be expressed as the problem of decomposing the $N\times C$ matrix $\mathbf{Y}$ of observed concentrations of $C$ pollutants into the product of two unknown matrices $\mathbf{G}$ and $\mathbf{H}$, of compatible dimensions, e.g. $N\times Q$ and $Q\times C$, representing, respectively, the emissions of $Q$ sources and their contributions to the $C$ pollutants' concentrations. 
Statistical solutions to this problem have been rooted in principal components analysis (specifically, the method of unmix, \cite{henry1997,henry2003,thurston1985,lewis2003}), factor analysis \citep{blifford1967} and latent variable models \citep{christensen2002,park2002b}. To distinguish the latter solutions from those based on deterministic approaches, statistical solutions to the source apportionment problem are referred to in the literature under the umbrella term of \emph{receptor models} or \emph{receptor modeling}.
As both matrices $\bm{G}$ and $\bm{H}$ in receptor models have to contain exclusively positive or, more generally, non-negative entries, one class of solutions focused on ways to impose the positivity condition on the matrices elements. Among these solutions the most famous is Positive Matrix Factorization (PMF), an algorithm introduced by \cite{paatero1994} and \cite{paatero1997} extremely popular in the late 1990's -2000 \citep{brown2007,kim2007}, partly also because of easy-to-use software developed by the US EPA \citep{norris2014}. 
In other factor analysis approaches, researchers placed most of their efforts in either determining conditions or methods that could ensure uniqueness of the solution \citep{koutrakis1987} or in identifying conditions that can guarantee identifiability \citep{park2002a}.
Recognizing that the literature on receptor models  is vast and dates back to the late 60's and 70's, we refer interested readers to \citet{hopke2016,pollice2011recent,krall2019} for extensive reviews. 
Here we focus mostly on reviewing Bayesian approaches to the source apportionment problem which sets themselves apart in that they have a principled and more encompassing way to handle uncertainty.

Bayesian approaches to source apportionment have been concentrated on several aspects: some work  focused on establishing priors for either the $\mathbf{G}$ or the $\bm{H}$ matrix or their elements \citep{billheimer2001,lingwall2008,park2004locating}, others developed model selection criteria to determine the number of components \citep{park2002b,park2004locating}, others focused on determining the conditions to guarantee identifiability of model parameters \citep{park2002a,park2002b}. An example is \citet{park2002b} who offer identifiability conditions in the case of multivariate pollutant data, with multi-dimensionality due either to the fact that the data refer to one pollutant measured at multiple sites or due to the fact that the data refer to multiple pollutants observed at a single site.

Some of the source apportionment models specified within a Bayesian framework are specified on the log scale, that is for the log concentration rather than the actual pollutants' concentrations, providing what is termed a source apportionment or receptor model with \emph{multiplicative error structure}, see \citet{nikolov2011,hackstadt2014}. In particular, \citet{nikolov2011} extend the previously proposed model by \citet{wolbers2005linear} to specify it in a Bayesian framework and to incorporate covariates' information, thus accounting for their effects on the source apportionment, whereas \citet{hackstadt2014} leverage information on the sources' emissions composition available from national database curated by EPA.
Other Bayesian source apportionment approaches were developed with the goal of explicitly accounting for temporal and/or spatial dependence in the pollutants' concentration data. Examples include: \citet{heaton2010} who, building upon the work of \citet{lingwall2008}, propose a dynamic linear receptor model for air pollution data collected over time, with the source contributions varying in time and provided with a Dirichlet prior; and \cite{park2001} who model multiple VOC species measured over time using a model where the source contributions to the VOC species are assumed unknown but constant in time while the source emission profiles are assumed to vary dynamically in time. On the other hand, spatial dependence is accounted for in the Bayesian multivariate receptor models proposed by \citet{jun2013multivariate,park2018,pollice2012major} who analyzed data relative to either multiple pollutants collected at multiple monitoring sites \citep{jun2013multivariate,park2018} or a single pollutant observed at multiple sites \citep{pollice2012major}.
Finally, recent work by \citet{tang2020} extends previous work by modeling the source emissions on a given day as depending on meteorological covariates, while \citet{park2015,park2016} develop  robust Bayesian multivariate receptor models by either assuming a heavy-tailed error distribution \citep{park2015} or by using a quantile regression approach \citep{park2016}.

Our work inserts itself in this literature by proposing a Bayesian solution to the source apportionment problem when the goal is to represent the concentrations of $C$ components of PM$_{2.5}$ observed over time as the additive result of unobserved sources’ emissions and sources' contributions, assuming that the number of sources is unknown and to be estimated. 
 
Our source apportionment model explicitly accounts for the spatial and temporal dependence in the pollutants' concentration data and it estimates the number of unknown sources by utilizing shrinkage priors on source-specific parameters following a strategy similar to that adopted by \cite{montagna2012} in a different context. Additionally, our model allows for characteristics of the monitoring locations to influence the level of local emissions, that is the amount of emissions generated by a source that is experienced at the monitoring location.  The model has been tailored for application to our motivating dataset, which refers to observations of six selected components of PM$_{2.5}$ at 32 monitoring sites in California during year 2021.

The article is structured as follows:
\Cref{sec:data} introduces the PM$_{2.5}$ speciation data that motivated our study; \Cref{sec:model} presents the Bayesian latent functional factor model that we propose for estimating the number of sources, the space-time varying emission profiles of the sources and the sources' contributions to the pollutants concentration. \Cref{sec:res} discusses the results obtained when applying the proposed model to both simulated data and to the observed speciated PM$_{2.5}$ concentrations in California, while \Cref{sec:discussion} concludes the paper with a discussion.

\section{Data}
\label{sec:data}
\subsection{PM$_{2.5}$ components} 
\label{subsec:pm25data}
Our motivating dataset refers to observations of selected components of PM$_{2.5}$ in California during 2021. Fine particulate matter, or simply PM$_{2.5}$, is a compound pollutant formed by particles of different types generated by a multitude of sources and processes, whose composition varies spatially and temporally. A crude classification of the components of PM$_{2.5}$ enlists the following types: water-soluble inorganic ions, carbon-containing components, inorganic elements and organic matter. In selecting which components of PM$_{2.5}$ to analyze in our study, we have kept in mind not only the relevance of the component in terms of the fraction of PM$_{2.5}$'s total mass they account for, but also their importance from an epidemiological perspective.

Water-soluble inorganic ions account for about 30 to 50\% of the total mass of PM$_{2.5}$ and include secondary ions such as sulfates (SO$_{4}^{2-}$) and nitrates (NO$_{3}^{-}$), which are the most abundant among the components of PM$_{2.5}$. Various epidemiological studies  have indicated that long-term exposure to sulfates and nitrates significantly increases the risk of adverse health events in humans, such as the risk of ischemic stroke \citep{ostro2007effects}, 
or the risk of cardiopulmonary crises in COPD patients \citep{zhou2021acute}, 
to name a few. Because of their relevance from a human health perspective, in our study we  have elected to use data relative to the concentrations of sulfate and nitrate.

Another important contribution to PM$_{2.5}$'s total concentration is provided by the carbon-containing components, which account for about 20 to 50\% of PM$_{2.5}$'s total mass. These components are mainly made up of organic carbon (OC), elemental carbon (EC), and carbonate carbon (CC), with the latter constituting only a small fraction of PM$_{2.5}$. Because OC and EC are the preponderant carbon-containing components of PM$_{2.5}$, we have decided to also obtain data on the concentration of OC and EC for 2021 in California.

Inorganic elements and organic matter constitute a smaller fraction of PM$_{2.5}$'s total mass. However, because of their ability to react with other elements to generate new compounds and because of their significant adverse health effects, we have decided to include also data relative to sulfur and aluminum in our spatio-temporal analysis. The first, sulfur, is present within PM$_{2.5}$ as part of sea aerosols, but it also contributes to PM$_{2,5}$'s total mass either in the form of sulfur dioxide (SO$_{2}$), mostly released during volcano emissions, or as a sulfur compound resulting from biogenic processes. On the other hand, aluminum, a metal commonly found on the Earth's crust, can be dispersed in the atmosphere from processes such as mining or by industrial facilities. Aluminum in PM$_{2.5}$ has been found to have serious adverse health effects, mostly of neurological type \citep{alasfar2021,vlasak2024}.

Limiting ourselves to these six PM$_{2.5}$ components, in this study, we set as a goal that of identifying the sources and processes that contributed to the concentrations of aluminum, sulfur, organic carbon, elemental carbon, nitrate, and sulfate across California in 2021. For this goal, we use observations of daily pollutants' concentrations collected during the year 2021 at monitoring sites within the Chemical Speciated Network (CSN) and the IMPROVE (Interagency Monitoring of Protected Visual Environments) networks, the latter a network of sites located in national parks and wilderness areas. We  have retrieved daily concentration data from the US Environmental Protection Agency outdoor air quality web portal \texttt{AirData} (available at \url{https://aqs.epa.gov/aqsweb/airdata/download_files.html}), downloading the pre-generated daily file of speciated PM$_{2.5}$ concentrations for the year 2021, retaining only the observations relative to California and the six components listed above.

\begin{figure}[!t]
    \centering
    \includegraphics[width=0.75\linewidth]{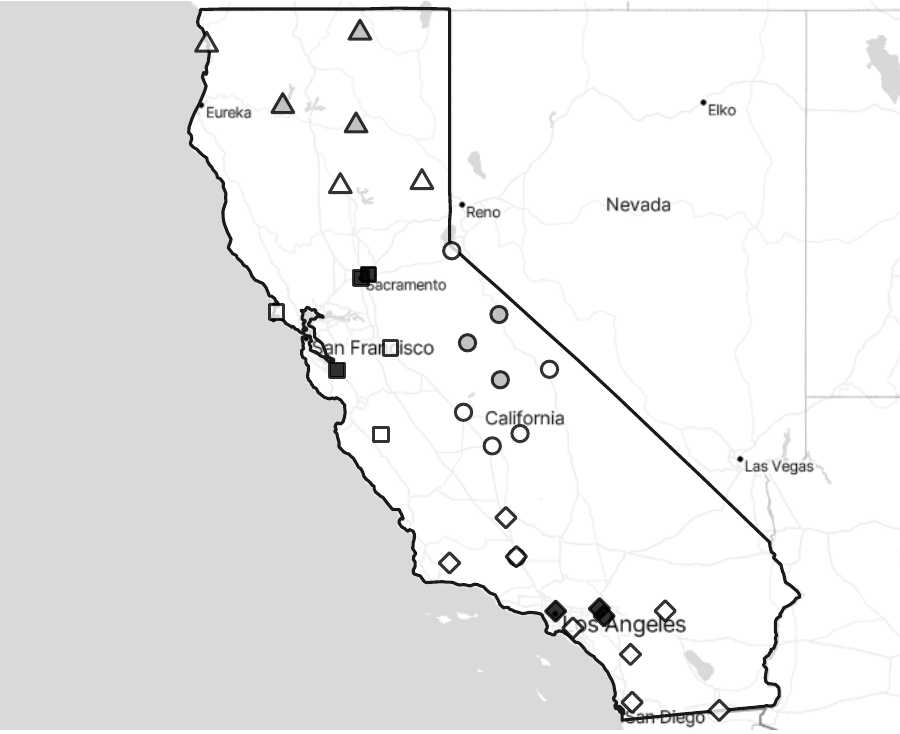}
    \caption{Locations of the $N=32$ monitoring sites reporting observations of concentrations of aluminum, sulfur, elemental carbon, organic carbon, nitrate, and sulfate in California in the year 2021. Different symbols (triangles, circles, squares, and diamonds) denote monitoring sites within the 4 subregions of California that we have created and denominated as North, Central-East, Central-West, and South, respectively. Filled symbols are used for the subset of monitoring sites within each of the four subregions, for which we show the pollutants' time series of log-concentrations in Figure~\ref{fig:eda_ts}.}
    \label{fig:cali_map}
\end{figure}

Figure~\ref{fig:cali_map} shows the locations of the 32 monitoring sites reporting observations of the six selected PM$_{2.5}$ constituents in the year 2021 in California. Sites are displayed with four different symbols, each representing one of the four different regions in which we partitioned California -- North, Central-East, Central-West, and South -- anticipating different pollution profiles during the course of the year across the four regions. As Figure~\ref{fig:cali_map} shows, monitoring sites are sparsely located across California, with some regions void of monitoring sites, while others display a great density. This is particularly true in and around large metropolitan centers, particularly in Southern California. Not all 32 stations report concentrations of all six PM$_{2.5}$ components, as Table~\ref{tab:eda_poll_nobs} indicates; organic carbon and elemental carbon are more scarcely measured than the remaining four pollutants. In general, monitoring sites within the two networks report levels of PM$_{2.5}$ species every three days, with some operating on a once-every-6-day schedule. Table~\ref{tab:eda_poll_nobs} provides a breakdown of the percentage of monitoring sites reporting observations every 3 days and every 6 days, respectively, along with some summary statistics regarding the number of observations per site.

\begin{table}[!b]
\caption{Number of monitoring sites reporting observations of a certain PM$\_{2.5}$ constituent's concentration, percentage of monitoring sites reporting observations every 3 days, respectively, every 6 days, and minimum - median - maximum number of observations per site.} \label{tab:eda_poll_nobs}%
\begin{tabular*}{\columnwidth}{@{\extracolsep\fill}c|cccc@{\extracolsep\fill}}
\toprule
 PM$_{2.5}$     & No. of monitoring  & Percent sites & Percent sites & Min-Median-Max\\
component & sites   & on a 1-in-3-day schedule & on a 1-in-6-day schedule & no. of obs. per site \\
\hline
  Aluminum & 32 & 75 \% & 25\% & 39 - 110 - 122 \\ 
  Sulfur &  32 & 75 \% & 25\% & 39 - 112 - 122 \\
  OC &  24 & 100\% & 0\% & 66 - 112 - 122 \\
  EC &  24 & 100\% & 0\% & 66 - 112 - 122 \\ 
  Nitrate &  32 & 75 \% & 25\% & 39 - 113 - 122 \\  
  Sulfate &  32 & 75 \% & 25\% & 39 - 113 - 122 \\
\hline
\end{tabular*}
\end{table}

As pollutant concentrations are non-negative, to avoid problems when generating predictions of PM$_{2.5}$ components' concentrations using a statistical model, it is common practice to model pollutants' concentrations on the log scale. Embracing common practice, we also work on the log scale, transforming predictions of the six PM$_{2.5}$ constituents' concentrations back to the original scale for ease of interpretation of the results. 

\begin{figure}[!t]
    \centering
    \includegraphics[width=\linewidth]{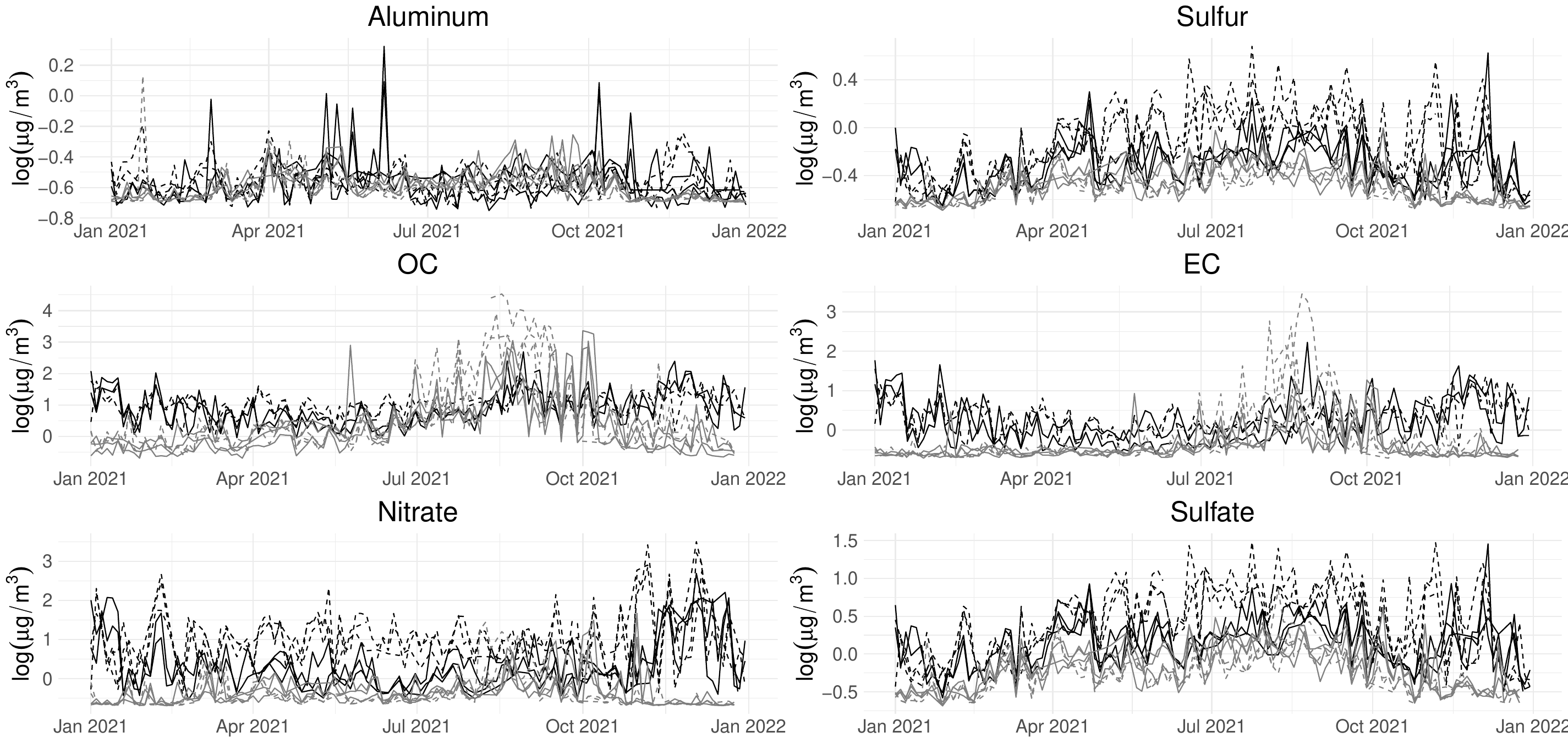}
    \caption{Time-series of the concentrations of the six pollutants under study during year 2021 at 12 different locations presented on the log scale. Time series displayed as black dashed (solid) lines refer to monitoring sites in the Southern (Central-West) area which are indicated by black diamonds (squares) in \Cref{fig:cali_map}. Time series displayed as gray dashed (solid) lines refer to concentrations measured at monitoring sites in the Northern (Central-East) area which are indicated by gray triangles (circles) in \Cref{fig:cali_map}.}
    \label{fig:eda_ts}
\end{figure}

Besides allowing us to circumvent problems due to the non-negativity of pollutants' concentrations, working on the log scale also addresses issues linked to the right-skewness of the distribution of PM$_{2.5}$ constituents' concentrations. In addition, it makes the measurements more comparable. This can be clearly seen in Figure~\ref{fig:eda_ts}; the figure shows the time series of log concentrations for the six PM$_{2.5}$ constituents at 12 randomly selected sites. The twelve selected sites  have been picked so that each of the four regions in California are represented by three sites. The locations of the three selected sites within each region can be recognized in Figure~\ref{fig:cali_map} by the fact that the corresponding symbol is filled in black or gray. To distinguish among the four regions, in Figure~\ref{fig:eda_ts}, profiles of pollutant concentrations relative to sites falling within the same region are displayed with the same plotting symbol. Specifically, dashed black lines are used for data corresponding to sites within the Southern region, whereas solid black lines are used to denote data relative to monitoring sites in the Central-West region. Conversely, dashed gray lines are employed for concentrations of pollutants observed at sites in the Northern region, whereas solid gray lines are reserved for sites in the Central-East region. Adopting these plotting conventions allows one to see more clearly patterns in the pollutants' concentrations that are more typically shared among sites within the same region but less so among sites belonging to different regions. For example, Figure~\ref{fig:eda_ts} shows that OC and EC have different trends in the North-East compared to the South-West. Moreover, Figure~\ref{fig:eda_ts} highlights seasonal differences in the concentration of the six PM$_{2.5}$ components. While sulfate and sulfur tend to have higher concentrations between April and October, in those months nitrates are less present in the air. On the other hand, the time series of OC and EC concentrations seem to display greater spatial and temporal variability among sites than the other pollutants. Specifically, while EC tend to have fairly low concentrations for most of the year in the North and Central-East regions with greater concentrations over the late summer months particularly in the North, in the South and Central-West, the concentration of EC tends to be higher in the winter months of November, December, January and February to slowly decrease and peak again around September. The same type of behavior over time and the same type of trends in the discrepancies among regions can be noticed in the panel displaying the time series of OC's concentrations. For aluminum and sulfur, the differences between the observed concentrations over time are less striking and seem to be more occasional than for the other pollutants discussed above.
The differences in the temporal behavior of the six PM$_{2.5}$ concentrations across the four regions clearly indicate the need to account for spatial dependence in our proposed model.

To further highlight the similarity in range and values among some pollutants' concentrations, particularly after the log transformation, Table~\ref{tab:eda_poll_summary} reports summary statistics (e.g. minimum, mean, maximum and variance) for the concentrations of each of the six PM$_{2.5}$ constituents on the log scale.

\begin{table}[!h]
\caption{Minimum, maximum, mean and variance of the observed concentrations of the six PM$_{2.5}$ components on the log scale.} \label{tab:eda_poll_summary}%
\begin{tabular*}{\columnwidth}{@{\extracolsep\fill}c|cccc @{\extracolsep\fill}}
\hline
   PM$_{2.5}$  & Minimum & Mean & Maximum & Variance \\ 
  component & (in $\log(\mu g/ m^3)$) &  (in $\log(\mu g/ m^3)$) &  (in $\log(\mu g/ m^3)$) & (in $(\log(\mu g/ m^3))^2$) \\ 
  \hline
  Aluminum & -0.78 & -0.56 & 0.32 & 0.02 \\ 
  Sulfur & -0.69 & -0.33 & 0.68 & 0.06 \\ 
  OC & -0.69 & 0.55 & 4.53 & 0.64 \\ 
  EC & -0.71 & -0.20 & 3.45 & 0.27 \\ 
  Nitrate & -0.70 & 0.11 & 3.50 & 0.55 \\ 
  Sulfate & -0.69 & 0.07 & 1.49 & 0.17 \\ 
  \hline
\end{tabular*}
\end{table}

\subsection{Covariates}
\label{subsec:covariates}

\begin{figure}[!t]
    \centering
    \includegraphics[width=.95\linewidth]{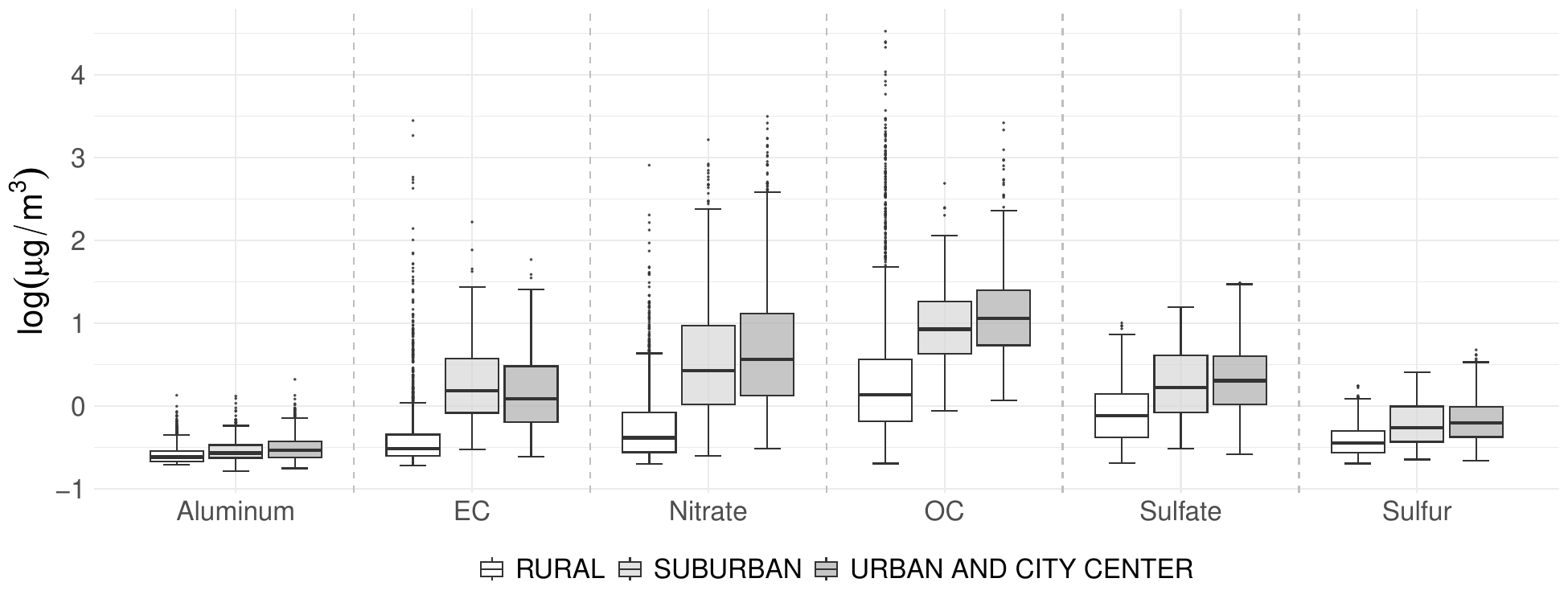}
    \caption{Boxplots displaying the distribution of concentrations,  on the log scale, for each of the six PM$_{2.5}$ components during year 2021 grouped by level of urbanization.}
    \label{fig:eda_boxplot}
\end{figure}

Besides measurements of air pollutants' concentrations, the EPA air quality web portal \texttt{AirData} also provides information on the area surrounding the monitoring site. More specifically, the website hosts a dataset that reports information on the geographical coordinates, the elevation (in meters above the sea level) of each monitoring site and the type of land setting for an area surrounding the monitoring sites. The possible land types are ``rural", ``suburban" and ``urban and city center". 
\Cref{fig:eda_boxplot} examines whether the hypothesis that pollution level varies by urbanization is valid by showcasing boxplots of of the log concentrations for the six PM$_{2.5}$ components' concentrations grouped by urbanization level. As the figure shows, in general, suburban and urban locations experience higher levels of air pollutants' concentrations compared to rural sites. This underscores the importance of including land use/land type as a covariate in our modeling framework. 

 On the other hand, exploratory analysis of a linear relationship between log concentration of the six PM$_{2.5}$ components and elevation did not reveal a clear trend: scatterplots investigating such relationship are very hard to read due to the repeated observations available at each site. However, since in general we expect higher levels of PM$_{2.5}$ mass at higher altitude sites, we include elevation of the monitoring site as a covariate in our source apportionment model.

\section{Model}
\label{sec:model}
We now introduce the model that we propose for the apportionment of the concentrations of the six PM$_{2.5}$ components across California during year 2021. Our interest is in developing a source-apportionment model for the continuous-space, discrete-time, multivariate stochastic process $\mathbf{Y}(\bm{s},t)=\left\{ Y_1(\bm{s},t), \ldots, Y_C(\bm{s},t) \right\}$ representing the log concentration of $C$ components of PM$_{2.5}$ as $\bm{s}$ varies in $\mathcal{S} \subset \mathbf{R}^2$ and $t$ varies in $\mathcal{T}\subset \mathbf{N}$. In our case, $\mathcal{S}$ denotes the two-dimensional subset of $\mathbf{R}^2$ representing the state of California while $\mathcal{T}$ denotes the set $\left\{ 1, 2, \ldots, 365 \right\}$ corresponding to days in year 2021. 

As the goal of a source apportionment model is to decompose the  concentration of a pollutant $c$ as the sum of the level of pollution generated by $K$ different sources, we express $y_c(\bm{s},t)$ as: 
\begin{equation}
\label{eq:fun_bsa}
    y_c(\bm{s},t) = \sum_{k=1}^{K} g_k(\bm{s},t)\ h_{k,c} + \epsilon_c(\bm{s},t), \quad 
    \epsilon_c(\bm{s},t) \iid \calN(0,\sigma_c^2) 
\end{equation}
for $c=1,\ldots,C$ and $t\in \mathcal{T}$. 
In \eqref{eq:fun_bsa}, $h_{k,c}$ represents the proportion of $g_k(\bm{s},t)$ that contributes to the log concentration of pollutant $c$ while $g_k(\bm{s},t)$ indicates the amount of pollution generated by source $k$ at time $t$ at location $\bm{s}$. In light of this, we call $h_{k,c}$ the \emph{contribution of 
source} $k$ to the concentration of pollutant $c$, whereas we refer to $g_k(\bm{s},t)$ as the \emph{local emissions} from source $k$ on day $t$ at location $\bm{s}$. This terminology breaks with previous work that commonly referred to the $h_{k,c}$'s as \emph{source composition profiles} or \emph{source (chemical) fingerprints} and called the $g_k(\bm{s},t)$'s as \emph{source contributions} on day $t$ at location $\bm{s}$ (see \citet{park2001,park2018,wolbers2005linear}). 
 
As both the $g_k(\bm{s},t)$'s and the $h_{k,c}$'s are unknown in our model, for sake of identifiability we cannot model them to be both spatially and temporally-varying. In choosing which of the two sets of parameters to allow to vary in space and time, we have elected
to keep the source contributions fixed in space and time for computational convenience as well as scientific suitability. Indeed, we believe that it is more plausible that the amount of emissions produced by a source varies spatially and temporally than it would its contribution to a pollutant's concentration. Additionally, as we will elaborate later, the $h_{k,c}$ terms must satisfy certain constraints that would be hard to enforce computationally if they were spatial processes.

Hence, we assume that the contributions $h_{k,c}$ do not change as $\bm{s}$ varies in $\mathcal{S}$ nor that they change during the course of the year.
Finally, we remark that since $y_c(\bm{s},t)$ denotes the log concentration of PM$_{2.5}$ component $c$ at location $\bm{s}$ on day $t$,  the decomposition in \eqref{eq:fun_bsa}, is an example of what is denominated as a \emph{source apportionment model with multiplicative error term} \citep{wolbers2005linear,park2018}.

In some of the previous solutions offered to the source apportionment problem, the number of sources $K$ was either assumed fixed and known \citep{park2002a}, leveraging domain-science knowledge, or estimated in a preliminary stage and then kept fixed in the source apportionment analysis \citep{jun2013multivariate}. In our work, we model the true number of sources as an unknown, \textit{random} quantity that we estimate from the data during our model fitting procedure. Regardless of whether the number of sources is considered known or unknown, a typical assumption of source apportionment models is that of \textit{mass balance}, that is, the assumption that, once created, pollution cannot be removed or eliminated. As a consequence, the source contributions $h_{k,c}$ are all non-negative numbers, for all $k$ and $c$.  
A further assumption is often imposed on the $h_{k,c}$'s: the sum of all the contributions corresponding to a given source $k$ over all the $C$ pollutants must be equal or less than 1 \citep{henry1991}. Under this assumption, it is possible that a source also contributes to the concentration of a pollutant $c^\prime$ not considered in the analysis. 
Here instead, we follow \cite{wolbers2005linear,lingwall2008,nikolov2011,hackstadt2014,park2018,rusanen2024} and we assume that the sum of all the contributions corresponding to a given source $k$ over all the $C$ pollutants must be equal to 1, e.g. $\sum_{c=1}^C h_{k,c} = 1$ for any $k$, thus postulating that each source contributes \emph{only} to the concentration of the $C$ components of PM$_{2.5}$ under study.
This constraint, introduced to guarantee identifiability, implies, as remarked by \cite{park2018}, that for each source $k$ only the relative contribution of each pollutant to the source's emissions can be estimated, and not the absolute amount.

To facilitate the estimation of the local emissions profiles $g_{k}(\bm{s},t)$, which account for the spatio-temporal dependence in the multivariate stochastic process $\bm{Y}(\bm{s},t)$, $\bm{s}\in \mathcal{S}$, $t\in \mathcal{T}$, we separate space and time, decomposing the \textit{local emissions from source $k$} at location $\bm{s}$ at time $t$, $g_k(\bm{s},t)$, into:
\begin{equation} 
   \ g_k(\bm{s},t)  = \gamma_{k}(\bm{s}) \ f_k(t),\  \ 
    \gamma_{k}(\bm{s})  =  \exp(\bm{X}(\bm{s}) \bm{\beta}_k + w_{k}(\bm{s})), \ \
    \bm{s}\in \mathcal{S}, t \in \mathcal{T}; \; k=1,\dots,K,
\label{eq:g_ik}
\end{equation}
with $f_k(t)$ representing the overall emissions generated by source $k$ over time. Thus, we refer to $f_k(t)$ as the \textit{global emission profile} relative to source $k$.
In \eqref{eq:g_ik}, the term $\gamma_{k}(\bm{s})$ either amplifies, if larger than 1, or attenuates, if smaller than 1, the level of pollution experienced at $\bm{s}$ due to emissions generated from source $k$ on day $t$. Expecting that this amplification, respectively, attenuation will be of similar magnitude in geographically close locations, and assuming that the characteristics of the location might play a role in the magnitude of the modification, we express $\gamma_k(\bm{s})$ as a function of $p$ site-specific covariates, $\mathbf{X}(\bm{s})$, $p$ regression coefficients $\boldsymbol{\beta}_k=\left( \beta_{1,k}, \beta_{2,k}, \ldots, \beta_{p,k} \right)^\prime$, and a spatial random effect term, $w_k(\bm{s})$. For each source $k$, the vector of regression coefficients $\boldsymbol{\beta}_k$ is assumed to follow a $p$-variate normal distribution with mean $\bm{m}_0$ and covariance matrix $s_0 \bm{I}_p$.
In turn, we place a $p$-variate normal distribution and an Inverse Gamma distribution, respectively, as prior on the parameters $\bm{m}_0$ and $s_0$. In other words, we have:
\begin{eqnarray}
\bm{\beta}_k | \bm{m}_0, s_0 \iid \mathcal{N}_p (\bm{m}_0, s_0 \bm{I}_p), \quad k=1,\ldots,K  \label{eq:prior_beta} \\
\bm{m}_0 \sim \mathcal{N}_p (\bm{0}, \bm{I}_p), \quad
s_0 \sim \text{InvGamma}(a_\beta,b_\beta). \nonumber
\end{eqnarray}
The spatial random effect $w_k(\bm{s})$, $\bm{s}\in \mathcal{S}$, is, in turn, modeled as a mean-zero Gaussian process; it accounts for the spatial correlation in the term $\gamma_k(\bm{s})$, capturing the spatial dependence in the local emissions source profiles, $g_k(\bm{s},t)$, that cannot be explained by the characteristics $\mathbf{X}(\bm{s})$ of the site alone. We take $w_k(\bm{s})$ to be source-specific as we expect that the attenuation, respectively, amplification in local emissions might vary from source to source. We model the spatial correlation in $w_{k}(\bm{s})$ using an exponential covariance function; this implies that, for a given $k=1,\ldots,K$, and any pair of points $\bm{s}$ and $\bm{s}^{\star} \in \mathcal{S}$,
\begin{equation}
\mbox{Cov}(w_k(\bm{s}),w_k(\bm{s}^{\star})) = \sigma^2_k \exp\left(- \frac{\| \bm{s}-\bm{s}^{\star} \|}{\rho_k}\right)
\label{eq:covwk}
\end{equation}
with $\| \bm{s} - \bm{s}^{\star} \|$ denoting the geographical distance between locations $\bm{s}$ and $\bm{s}^{\star}$. In \eqref{eq:covwk}, $\sigma^2_k$ indicates the marginal variance of the spatial random effects, and $\rho_k$ denotes the range parameter which summarizes the scale of the spatial dependence in the spatial process $w_{k}(\bm{s}), \bm{s}\in \mathcal{S}$. Assuming that there might be different scales of spatial dependence in the local emissions generated by various sources, we allow each spatial random effect $w_k(\bm{s}),\bm{s}\in \mathcal{S}$, to have its own range parameter $\rho_k$. However, we provide all range parameters with the same prior, an Inverse Gamma distribution. Thus: $\rho_k \iid \text{InvGamma}(a_\rho,b_\rho)$ for $k=1,\ldots, K.$ On the other hand, since the marginal variance of the spatial random effects might not be identifiable, we set $\sigma^2_k \equiv 1$, for all $k=1,\ldots, K$. 

Considering again \eqref{eq:g_ik}, we justify our choice of an exponential link function to relate $\gamma_k(\bm{s})$ to the covariates $\mathbf{X}(\bm{s})$ and the spatial random effect $w_k(\bm{s})$ by the intention to guarantee that the global and the local emissions profiles relative to source $k$ have the same sign at every location $\bm{s}\in \mathcal{S}$. A discrepancy in the sign of the two profiles for a given source could indicate differences in the temporal evolution of the source's emissions that would be difficult to reconcile. As an example, global peaks could become drops when considered locally and vice versa, suggesting different types of seasonality in the emissions associated with the same source. 

Moving onto the global emission profile $f_k(t)$ associated with source $k$, this terms accounts for the temporal correlation in the continuous-space, discrete-time multivariate stochastic process $\mathbf{Y}(\bm{s},t)$, $\bm{s}\in \mathcal{S}$, $t\in \mathcal{T}$. Rather than modeling the correlation using an autoregressive or a dynamic model specification, we adopt a functional approach and express $f_k(t)$
as a linear combination of basis functions with source-specific basis function coefficients. Taking as basis functions $M$ cubic B-splines $\{b_m(t)\}_{m=1}^{M}$,  we have: 
\begin{equation} 
\label{eq:f_k}
    f_k(t)= \sum_{m=1}^M \lambda_{k,m}b_m(t), \qquad k=1,\dots,K
\end{equation}
where the number $M$ of basis functions is fixed a priori and taken to be large.

To determine the \emph{true} number $K^{\star}$ of sources whose global emissions profiles contribute, through the corresponding local source profiles, to the observed concentration of $C$ pollutants, we impose sparsity on the basis function coefficients $\lambda_{k,m}$ via a careful prior specification. Indeed, sparse priors are often
employed in latent factor models to avoid overparametrization. The sparsity-inducing prior will also yield, as a byproduct, an estimate of the \textit{true} number $K^*$ ($K^*\leq K$) of sources. 
More specifically, following \cite{montagna2012}, we specify a \textit{multiplicative gamma process shrinkage} (MGPS) \citep{bhattacharya2011} prior on the coefficients $\lambda_{k,m}$ of the cubic B-splines in \eqref{eq:f_k}: for $k=1,\dots,K$, and $m=1,\dots,M$
\begin{equation}
\begin{split}
\label{eq:mgps}
    \lambda_{k,m}| &\phi_{k,m}, \eta_k \ind \mathcal{N}\Big(0,\ \phi_{k,m}^{-1} \eta_k^{-1}\Big),  \quad \phi_{km} \iid \text{Gamma}\Bigg(\frac{\nu}{2}, \frac{\nu}{2}\Bigg)  \\
    \eta_k &= \prod_{l=1}^k \delta_l\ , \quad
 \delta_1 \sim \text{Gamma}\big(a_1, 1\big),  
    \ \delta_l \iid \text{Gamma}\big(a_2, 1\big),
   \ l \ge 2, \textrm{ with } a_1,a_2>1.
    \end{split}
\end{equation}
We will use the shorthand notation $\mbox{MGPS}(\nu,a_1,a_2)$ to denote the prior on the $\lambda_{k,m}$'s shown above in \eqref{eq:mgps}.
As in \cite{montagna2012}, we note that the $\eta_k$'s are global shrinkage parameters for the global emission profiles $f_k(t)$'s whereas the $\phi_{km}$'s are local shrinkage parameters for the $\lambda_{k,m}$'s.
Under the choice $a_2 > 1$, the $\eta_k$'s are stochastically increasing
favoring more shrinkage as $k$ increases. The choice of this shrinkage prior allows many of the $\lambda_{k,m}$'s to be close to zero, giving us the possibility to discard redundant sources. With this specification, the number of global sources, $K$ is an unknown quantity whose value is determined during model fitting. See Section~\ref{sec:infer} for more details in this regard.

For each source $k$, the vector of length $C$, $\mathbf{h}_k:=(h_{k,1},\ldots,h_{k,C})^\prime$, denotes the contribution of source $k$ to the concentrations of pollutants $1, 2,\ldots, C$. In specifying a prior for this vector, since we postulate that $\sum_{c=1}^{C} h_{k,c}=1$, we assume that 
\begin{equation}
\label{eq:prior_h}
    \bm{h}_k \iid \text{Dir}(\bm{\alpha}_0) ,  \quad k=1,\dots,K.
\end{equation}
In \eqref{eq:prior_h}, $\text{Dir}(\bm{\alpha}_0)$ denotes the $C$-dimensional Dirichlet distribution with parameter $\boldsymbol{\alpha}_0 = (\alpha_0, \dots, \alpha_0)$, with $\alpha_0>0$.
Alternatively, other marginal priors for $\bm{h}_k$ could be specified, as discussed in \cite{hackstadt2014,lingwall2008,nikolov2007}.

Finally, we conclude the prior specifications for our model, by stating our choice for the marginal prior distribution of the pollutant-specific standard deviation $\sigma_c$:
\begin{equation}
\label{eq:prior_epsilon}
    \sigma_c \iid \text{Cauchy}^+(0,1)\ , \qquad 
    c=1,\dots,C
\end{equation}
where $\text{Cauchy}^+(0,1)$ denotes the half standard Cauchy distribution. This prior distribution was selected following the suggestion in \cite{gelman2006}.  We implicitly assume a priori independence among blocks of parameters for which we have specified the marginal prior distributions. 

As it appears from \eqref{eq:fun_bsa}-\eqref{eq:g_ik}, our model specification allows us 
to borrow information across the three different levels of data -- space, time and different pollutants --
to obtain precise estimates of the parameters of interest: the \textit{true} number of sources $K^*$, the local source emission profiles, the $g_k(\bm{s},t)$'s, and the sources' contributions to the pollutants concentrations, the $h_{k,c}$'s. At the same time, our model also suffers from unidentifiability: for example, we are unable to uniquely separate and estimate the $\gamma_k(\bm{s})$'s and the $f_k(t)$'s, 
and, as a consequence, we cannot uniquely identify the regression coefficients $\boldsymbol{\beta}_k$ nor the spatial random effects $w_k(\bm{s})$, corresponding to each source.
Nonetheless, we want to reiterate that, in the context of our model's inferential goal, these can be considered as nuisance parameters; additionally, despite our inability to precisely estimate these parameters, we are still able to draw some conclusions on them. For example, we 
can estimate if, for a given source, the effect of a covariate on the local source emissions profile
is positive or negative, or we can compare the relevance of the regression parameters across sources.

We conclude this section by reformulating our model in a matrix form, which allows us to highlight its hierarchical structure as well as its connection to factor models, one of the statistical approaches historically employed for source apportionment. The matrix formulation is also extremely helpful when developing MCMC sampling algorithms.

Having observed data at $N$ monitoring sites, $\bm{s}_1, \bm{s}_2, \ldots, \bm{s}_N$, we stack the observations $y_{c}(\bm{s}_i,t_{ij})$ of log concentrations of PM$_{2.5}$ components, $c=1,\ldots,C$, taken at monitoring site $\bm{s}_i$ on day $t_{ij}$, with $t_{ij} \in \mathcal{T}_i \subset \mathcal{T}$, to form the $\left( C\cdot L \right) 
\times 1$ vector $\mathbf{Y}$ with $L$ defined as $L=\sum_{i=1}^{N} l_i$ and $l_i=| \mathcal{T}_i|$, number of days in year 2021 during which the monitor at location $\bm{s}_i$ measured air pollutants concentrations. We note that this notation implicitly assumes that the level of all $C$ pollutants are measured on the days in which a monitoring site is operational which might not always be the case. In this eventuality, the value of the log concentration is set to missing and treated as an additional parameter to be estimated during model fitting.
The vector $\mathbf{Y}$ is organized so that $\mathbf{Y}=\left( \mathbf{Y}_1, \ldots, \mathbf{Y}_N \right)^\prime$ with $\mathbf{Y}_i=\left( y_1(\bm{s}_i, t_{i1}), y_2(\bm{s}_i, t_{i1}), \ldots, y_C(\bm{s}_i, t_{il_{i}}) \right)^\prime$ a $(C\cdot l_i)$-dimensional vector for $i=1,\ldots,N$. 
Denoting with $\bm{G}_i$ the $l_i \times K$ matrix whose $j$-th row, $j=1,\ldots,l_i$, is the $K$-dimensional row vector $\left( g_1(\bm{s}_i,t_{ij}), g_2(\bm{s}_i,t_{ij}),\ldots, g_K(\bm{s}_i,t_{ij}) \right)$ and indicating with $\bm{H}$ the $K\times C$ matrix row-stochastic matrix whose $m$-th column, $m=1,\ldots,C$, is the $K$-dimensional vector $\left( h_{1,m}, h_{2,m}, \ldots, h_{K,m} \right)^{\prime}$, from \eqref{eq:fun_bsa} it follows that:
\begin{equation}
\mathbf{Y}_i=\mbox{vec}\left( (\bm{G}_i \cdot \bm{H})^\prime \right) + \boldsymbol{\epsilon_i} = \mbox{vec} \left( \bm{H}^\prime \cdot \bm{G}_i^{\prime} \right) + \boldsymbol{\epsilon_i} 
\label{eq:yi_vec}
\end{equation}
where $\boldsymbol{\epsilon_i}$ is the $(C\cdot l_i)$-dimensional vector $\boldsymbol{\epsilon_i}=\left( \epsilon_1(\bm{s}_i,t_{i1}),\epsilon_2(\bm{s}_i, t_{i1}), \ldots, \epsilon_C(\bm{s}_i, t_{il_{i}}) \right)^\prime$ and \textit{vec} is the \textit{vectorization} operation that transforms matrices into vectors by concatenating the matrix's columns. In other words, the result of $\mbox{vec}(\bm{H}^\prime \cdot \bm{G}^\prime_i )$ is the $C\cdot l_i$-dimensional vector $\left( \sum_{k=1}^{K} g_k(\bm{s}_i,t_{i1}) h_{k,1}, \sum_{k=1}^{K} g_k(\bm{s}_i,t_{i1}) h_{k,2}, \ldots, \sum_{k=1}^{K} g_k(\bm{s}_i,t_{il_{i}}) h_{k,C} \right)^\prime$.
In turn, from \eqref{eq:g_ik}, the $l_i \times K$ matrix $\bm{G}_i$ can be written as:
\begin{equation}
\bm{G}_i=\bm{F}_i \cdot \boldsymbol{\Gamma}_i
\label{eq:g_matrix}
\end{equation}
where $\bm{F}_i$ is a $l_i \times K$ matrix with $j$-th row equal to the $K$-dimensional vector $\left( f_1(t_{ij}),f_2(t_{ij}),\right.$ $\left.\ldots, f_K(t_{ij}) \right)$ while $\boldsymbol{\Gamma}_i$ is the $K\times K$ diagonal matrix with main diagonal equal to $\left( \gamma_1(\bm{s}_i),\gamma_2(\bm{s}_i),\ldots,\gamma_K(\bm{s}_i) \right)$, thus implying that $\bm{G}_i^\prime=\boldsymbol{\Gamma}_i \cdot \bm{F}_i^\prime$. 
At the same time, from \eqref{eq:g_ik} it follows that the matrix $\boldsymbol{\Gamma}_i$ is given by:
\begin{equation}
\boldsymbol{\Gamma}_i = \mbox{diag} \left( \exp \left[ \bm{X}_i \boldsymbol{\beta} + \bm{w}_i \right] \right) 
\label{eq:gamma_matrix}
\end{equation}
where $\bm{X}_i$ is the $1\times p$ vector with entries equal to the values of the covariates at location $\bm{s}_i$, $\boldsymbol{\beta}$ is the $p\times K$ matrix of regression coefficients with $j$-th column the $p$-dimensional vector $\left( \beta_{j,1}, \beta_{j,2}, \ldots, \beta_{j,p} \right)^\prime$, $\bm{w}_i$ is the $1\times K$ vector $\left( w_{1}(\bm{s}_i), w_{2}(\bm{s}_i), \ldots, w_{K}(\bm{s}_i) \right)$, and \textit{diag($\bm{v}$)} is the operation that transform a $1\times K$ vector into a $K\times K$ diagonal matrix with main diagonal given by the vector $\bm{v}$.
On the other hand, \eqref{eq:f_k} implies that that the $K\times l_i$ matrix $\bm{F}_i^\prime$ can be written as:
\begin{equation}
\bm{F}_i^{\prime} = \boldsymbol{\Lambda} \bm{B}_i
\label{eq:f_matrix}
\end{equation}
with $\boldsymbol{\Lambda}$ the $K\times M$ matrix with $j$-th row given by the $M$-dimensional row vector $\left( \lambda_{j,1}, \lambda_{j,2}, \ldots \lambda_{j,M} \right)$ whereas $\bm{B}_i$ is the $M\times l_i$ matrix of $M$ cubic B-splines evaluated at times $t_{ij} \in \mathcal{T}_i$, e.g. the $j$-th column of $\bm{B}_i$ is the $M$-dimensional vector $\left( b_{1}(t_{ij}), b_{2}(t_{ij}), \ldots, b_{M}(t_{ij}) \right)^\prime$.
Combining \eqref{eq:yi_vec}, \eqref{eq:g_matrix}, \eqref{eq:gamma_matrix}, and \eqref{eq:f_matrix} together, we have that
\begin{equation*}
\bm{Y}_i = \mbox{vec} \left(\bm{H}^\prime \cdot \mbox{diag}\left( \exp \left[ \bm{X}_i \boldsymbol{\beta} + \bm{w}_i \right] \right) \cdot \boldsymbol{\Lambda} \bm{B}_i \right) + \boldsymbol{\epsilon}_i \qquad i=1,\ldots, N.
\label{eq:y_matrix}
\end{equation*}

Having expressed the model in matrix form, the hierarchical formulation of our model is as follows: for each $i=1,\ldots,N$,
\begin{eqnarray}
\label{eq:BayesianModel}
\bm{Y}_i | \bm{H},\bm{X}_i, \boldsymbol{\beta},\bm{w}_i,\boldsymbol{\Lambda},\bm{B}_i,\boldsymbol{\Sigma}_Y  & \ind & N_{C\cdot l_i}\left( \mbox{vec} \left( \bm{H}^\prime \cdot \mbox{diag}\left( \exp \left[ \bm{X}_i \boldsymbol{\beta} + \bm{w}_i \right] \right) \cdot \boldsymbol{\Lambda}\bm{B}_i \right), \boldsymbol{I}_{l_{i}}\otimes \boldsymbol{\Sigma}_Y \right) \notag\\
\bm{H} & = & \left( \bm{h}_1, \ldots, \bm{h}_K \right)^\prime \notag\\ 
\bm{h}_k & \stackrel{iid}{\sim} & \mbox{Dir}(\boldsymbol{\alpha}_0) \qquad k=1,\ldots, K \notag \\
\boldsymbol{\beta} | \bm{m}_0, s_0 & \sim & MN_{p\cdot K}(\bm{1}_K \otimes \bm{m}_0, s_0 \bm{I}_{p}, \bm{I}_K) \notag\\
\bm{W} & = & (\bm{w}^{\prime}_1, \bm{w}^{\prime}_2,\ldots,\bm{w}^{\prime}_N )\\
\bm{W}_{k \cdot} | \rho_k & \stackrel{ind}{\sim} & N_{N}(\mathbf{0}_N, \boldsymbol{\Sigma}_{k} ) \qquad k=1,\ldots,K \notag  \\
\boldsymbol{\Lambda} & = & (\lambda_{k,m})_{k=1,\ldots,K; m=1,\ldots, M} \notag \\
\lambda_{k,m} | \nu,a_1,a_2 & \sim & MGPS(\nu,a_1,a_2) \qquad k=1,\ldots, K; m=1,\ldots, M \notag
\end{eqnarray}
where $MN_{p\cdot K}$ denotes the Matrix Normal distribution of dimension $p\cdot K$, $\boldsymbol{\Sigma}_Y$ denotes the $C\times C$ diagonal matrix with elements $\sigma^2_1, \sigma^2_2, \ldots, \sigma^2_C$, $\bm{W}_{k \cdot}$ represents the k-th row of the $K\times N$ matrix $\bm{W}$, and $\boldsymbol{\Sigma}_{k}$ is a $N\times N$ matrix with $(j,m)$-th element equal to $\exp \left( -\frac{\| \bm{s}_j - \bm{s}_m \|}{\rho_k} \right)$.

\section{Posterior inference}
\label{sec:infer}
We fit our model \eqref{eq:BayesianModel} using the statistical software Stan (\cite{stan}) which uses an Hamiltonian Monte Carlo (HMC) algorithm and its adaptive variant, the no-U-turn sampler (NUTS) to generate posterior samples for each model parameter. As the no-U-turn sampler cannot provide posterior samples for parameters that assume discrete values, we have devised a specific model fitting procedure to determine the true number $K^\star$ of sources, that we explain in detail below.

In fitting the proposed source apportionment model to data, both simulated and real pollution data, we make the following choice for the hyperparameters. 
Following \cite{montagna2012}, in the MGPS prior for the cubic B-splines coefficients $\bm{\lambda_{k,m}}$ in \eqref{eq:mgps}, we  set $\nu=3$, and $a_1=10,\ a_2=20$, respectively, in order to encourage shrinkage. As proved in \cite{bhattacharya2011}, when $a_2>1$,  the $\eta_k$'s in \eqref{eq:mgps} are stochastically increasing, leading to greater shrinkage as the source index $k$ increases; see also \cite{durante2017note}. With such a selection of the hyperparameters in the shrinkage prior, many $\bm{\lambda}_{k,m}$'s will result close to zero while avoiding factor splitting, thus inducing an effective selection on the number of  sources. 
For the range parameters $\rho_k$ of the spatial random effects $w_{k}(\bm{s}), \bm{s}\in \mathcal{S}$, that contribute to the modification of the global emissions from source $k$ into the local source emissions, we set $a_\rho=3$ and $b_\rho=1000$. With this hyperparameter selection, each $\rho_k$ has a prior mean of $500\ km$ and large prior variability. We have elected these values for the hyperparameters keeping in mind the notorious issue of non-identifiability discussed by \cite{banerjee2014} in Section 6.4.3.3, and \cite{goicoa2018}, and following \cite{sahu2022} who suggests to choose hyperparameters so that the prior mean of the range parameter is approximately equal to one third of the maximum distance between sites (here $1248\ km$). Finally, we set the parameter of the Dirichlet distribution used for the source contributions $h_{k,c}$, presented in \eqref{eq:prior_h}, equal to $\alpha_0=1$.

We use an initial portion of the MCMC algorithm to estimate the unknown number of sources $K^*$. Once $K^{*}$ is determined, we use its estimated value
in the remainder of the MCMC iterations.
To obtain an estimate of $K^*$, we start the algorithm using $K=10$. We run the MCMC algorithm, updating and sampling all parameters for a first batch of 500 iterations.
Following \cite{bhattacharya2011}, at iteration 501, we count the number of rows $K_{curr}^*$ of $\boldsymbol{\Lambda}$ having all $M$ elements over a significance threshold $\zeta$ and we set it as the current estimate for $K^*$. We use $K^{\star}_{curr}$ as the new initial value for $K$, that is, we set $K=K_{curr}^*$. We repeat the same procedure with batches of 500 iterations until the value of $K$ and the estimate $K_{curr}^*$ of $K^{\star}$ are equal.
When the value of $K$ is not updated anymore, it means that no additional shrinkage can be provided by the algorithm. This procedure is part of the warming-up phase.
Posterior inference for model parameters is based on posterior samples generated after the warming-up phase and obtained by running one chain of the MCMC algorithm for 10,000 iterations. Of these 10,000 iterations, 6,000 were discarded for burn-in, with final 4,000 samples retained and used for posterior inference.

\section{Results}
\label{sec:res}
We now present results obtained by applying our source apportionment model to both simulated data and observations of PM$_{2.5}$ components' concentrations in California in 2021. In both cases, results refer to application of our model with prior specifications and hyperparameters as described in Section~\ref{sec:infer}.

\subsection{Simulation study}
\label{subsec:simulation}
To determine whether our model is capable to recover the true number of sources, their global and local emission profiles, as well as the sources' contribution to the observed pollutants' concentrations, we design a simulation study where we know all these quantities. To ensure that our simulated data are realistic, we use the locations of the speciated PM$_{2.5}$ monitors in California ($N=32$) as well as the monitoring sites' elevation and land use information (urban and city center, rural and suburban) as  covariates. We simulate concentration data for $C=6$ pollutants according to our model, using as true number of sources $K^{\star}=2$ with the two sources global emission profiles, $f_1(t)$ and $f_2(t)$, $t\in \mathcal{T}=\left\{ 1,2, \ldots, 365 \right\}$, generated using $M=15$ cubic B-splines and coefficients $\lambda_{k,m}$, $k=1,2$, $m=1,\ldots,15$ sampled from a standard normal distribution. For each source $k$, we simulate the local emission profiles, $g_{k}(\bm{s}_i,t)$, $i=1,\ldots,32$, following \eqref{eq:g_ik} with $\bm{\beta}_1=(-0.2, -0.5, -0.4, -0.2)^{\prime}$ and $\bm{\beta}_2=(0.4, 0.1,  0.2, -0.3)^{\prime}$, respectively, where the first entry represents the
intercept, and the last three entries represent the coefficients of suburban area indicator, urban and city center area indicators and elevation, respectively. The spatial random effects $w_1(\bm{s})$ and $w_2(\bm{s})$,  have been simulated at the 32 locations by generating, respectively, a sample from the 32-dimensional multivariate normal distribution with mean zero and covariance matrix induced by an exponential covariance function with marginal variance equal to $1$ and range parameters $\rho_1=600$ and $\rho_2=300$, respectively. Finally, the contributions of the two sources to the six PM$_{2.5}$ components' concentrations have been set equal to $\bm{h}_1 = (0.348, 0.086, 0.027, 0.028, 0.257, 0.254)$ and $\bm{h}_2 = (0.107, 0.290, 0.295, 0.025, 0.249, 0.034)$. The two vectors  have been obtained by sampling from the Dirichlet distribution with parameter $\alpha_0=1$.

\begin{figure} 
    \centering
    \includegraphics[width=\linewidth]{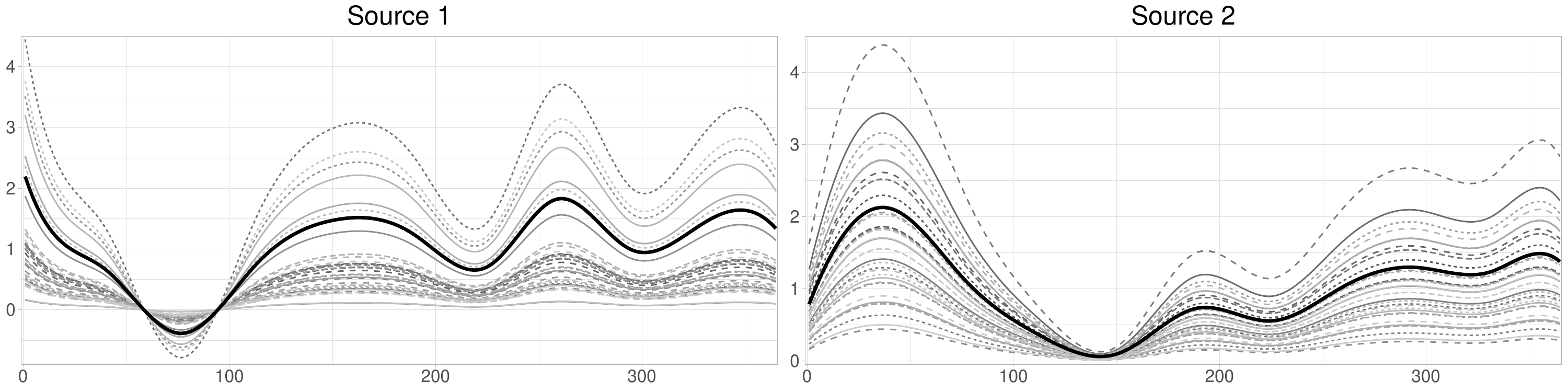}
    \caption{Simulated local emission profiles $g_k(\bm{s}_i,t)$, $i=1,\ldots,N$, $t\in \mathcal{T}$,  from the two true sources. 
    In both panel, the 32 local emission profiles are displayed with different plotting symbols in gray while the global source profiles, $f_1(t)$ and $f_2(t)$, are denoted with black solid lines.}
\label{fig:sim_fg}
\end{figure}

Having obtained the global and local source emissions profiles, we generate synthetic data $y_c({\bm s}_i,t)$, $i=1,\ldots,N$, $c=1,\ldots,C$, $t\in\mathcal{T}$, according to \eqref{eq:fun_bsa},  with $\sigma^2_c$ set equal to, respectively, 0.1, 0.2, 0.3 and 0.5 times the empirical variance of the simulated $\sum_{k=1}^{K} g_k(\bm{s}_i,t)\cdot h_{k,c}$, $i=1,\ldots,N$, $t\in\mathcal{T}$, for each $c=1,\ldots, C$. We consider different scenarios for $\sigma^2_c$ in order to establish how the ability of our model to correctly estimate the model parameters of interest deteriorates as the variance of the white noise in the data increases.
\Cref{fig:sim_fg} shows the simulated local source emission profiles relative to the two sources; solid lines represent the global source emission profiles $f_1(t)$ and $f_2(t)$.
As we have discussed in Section~\ref{sec:data}, observations of PM$_{2.5}$ components' concentrations are not available every day, thus to create simulated data that are more representative of a real air pollution dataset, for each site $i=1,\dots,N$, we discard many of the simulated $y_c({\bm s}_i,t)$ values, with $c=1,\ldots,6$ retaining only a random number $l_i$. As the majority of the EPA-maintained PM$_{2.5}$ species monitors operate on a 1-in-3-day schedule, as shown in Table~\ref{tab:eda_poll_nobs}, for each site $\bm{s}_i$, we draw the total number of observations $l_i$ in the year from a discrete uniform distribution $U\{100,101,\ldots, 150\}$. 
Having determined the total number of observations $l_i$ for each site, we have identified the simulated observations to retain at location $\bm{s}_i$ by sampling, for each pollutant $c$, $l_i$ values without replacement from the set $\left\{ y_c({\bm s}_i,t), t\in \mathcal{T} \right\}$.
We fit our source apportionment model \eqref{eq:fun_bsa}-\eqref{eq:prior_epsilon} to the synthetic data. Our goal is to evaluate whether we can obtain precise estimates of the true number of sources $K^*$, the local source emission profiles $g_k(\bm{s}_i,t)$, $i=1,\ldots,N$, $t\in \mathcal{T}$, and the sources contributions $h_{k,c}$, $c=1,\dots,C$ and $k=1,\dots,K^*$. 
\begin{figure} 
    \centering
    \includegraphics[width=.95\linewidth]{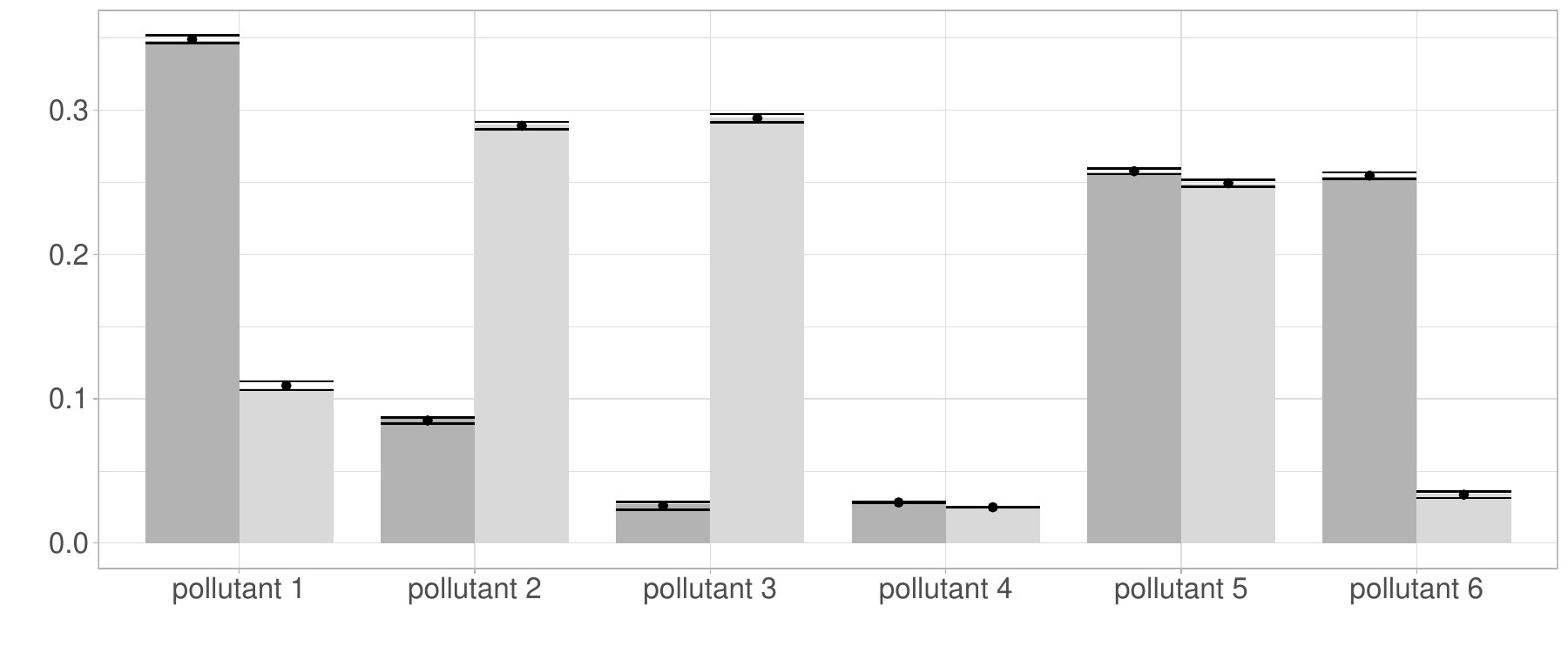}
    \caption{Posterior median (black dot) and 95\% credible interval (error bars) of $h_{k,c}$ for $k=1,2$ and $c=1,\dots,6$. The height of the bars represents the true values of the $h_{k,c}$'s: the contributions of source 1 to the six pollutant concentrations are denoted in darker gray,
    while contributions of source 2 are depicted in lighter gray.}
\label{fig:sim_postH}
\end{figure}

Examining the results for the case $\sigma^2_c=0.1$ of the empirical variance, we can see that, because of the effective shrinkage induced by the MGPS prior in \eqref{eq:mgps}, our model is able to recover the true number of sources $K^*=2$. The model yields also precise estimates of each source's contribution to the various pollutants concentration as \Cref{fig:sim_postH} shows. For each pollutant $c$ and each source $k$, the point estimates of the $h_{k,c}$'s, that here we take to be the posterior medians, are very close to the true values with 95\% credible intervals that are extremely narrow and always include the true value of the $h_{k,c}$'s for all pollutants and sources.

\begin{figure} \centering
    \includegraphics[width=\linewidth]{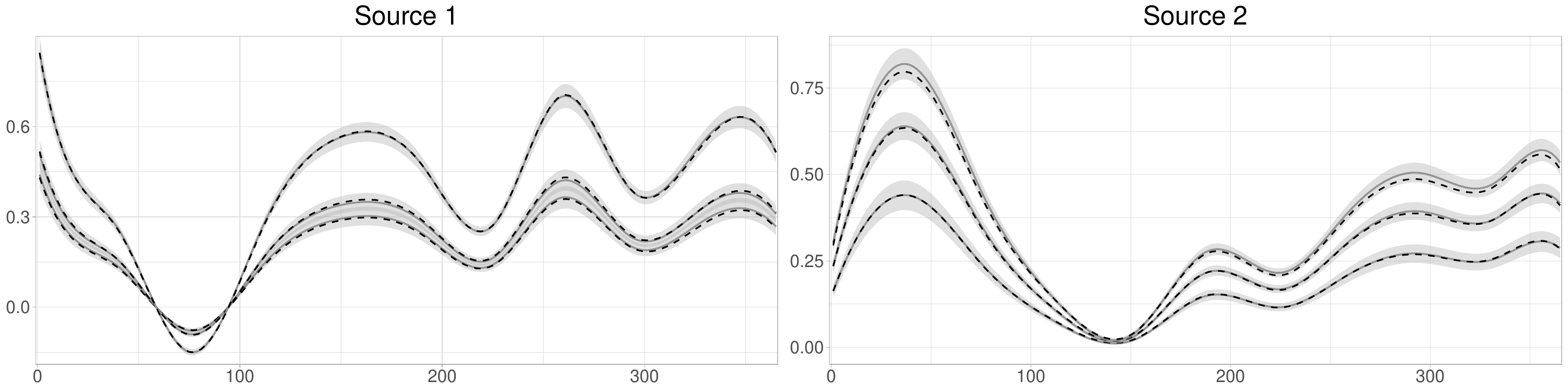}
    \caption{Posterior median (solid line) and 95\% pointwise credible intervals (gray shaded areas) for $g_k(\bm{s}_i,t)$, $t\in \mathcal{T}$, $k=1,2$, at three representative sites $i=9,11,18$ among the $N=32$ under study. In each panel, the black dashed lines represent the true (simulated) local emission profiles.}
    \label{fig:sim_postG}
\end{figure}

Our model also yields good estimates of the local emission profiles $g_k(\bm{s}_i,t)$ as \Cref{fig:sim_postG} illustrates. For sake of readability, for each source, the figure shows the estimated local emission profiles at three selected locations along with pointwise 95\% credible bands. For each location $\bm{s}_i$, the estimate of the local emission profile is taken to be the curve obtained by joining the posterior median of $g_k(\bm{s}_i,t)$ at each $t\in \mathcal{T}$. For the three locations considered in \Cref{fig:sim_postG}, we present the true local emission profile using dashed line while we display the estimated profile with a solid line, highlighting the nearly perfect agreement among the curves at all 3 locations. Analogous results hold for the other 29 locations not presented here.

Even though our model suffers from the issue of unidentifiability, 
the true values of the spatial ranges $\rho_1=600$ and $\rho_2=300$, respectively, lie within the 95\% credible intervals, which are $[327.19; 1056.01]$ and, $[124.29; 365.23]$, respectively.
Similarly, even though the estimates of the vectors of regression coefficients $\boldsymbol{\beta}_1$ and $\boldsymbol{\beta}_2$ are characterized by great uncertainty, as expected given the unidentifiability issue mentioned above and discussed in Section~\ref{sec:model}, as for the range parameters $\rho_1$ and $\rho_2$, the  pointwise 95\% credible intervals contain the true values (see Appendix). 
Moreover, the posterior medians and the true values of all the regression coefficients have the same sign.

The Appendix contains posterior estimates of the local emission profiles $g_k({\bm s}_i, t)$’s and the source contributions $h_{k,c}$'s for $\sigma^2_c$
equal to 0.2, 0.3 and 0.5 of the empirical
data variance. Here, we simply report that our model is able to correctly identify the true number of sources, $K^*=2$, even when there is increasing noise in the data.
This simulation study suggests that, despite the unavoidable non-identifiability that affects some parameters in our model, by fitting the proposed source apportionment model, we are able to estimate the local and global emission profiles, the source contributions to each pollutant's concentration, and the true number of sources $K^{\star}$ quite reliably.

\subsection{PM$_{2.5}$ species in California}
\label{subsec:PM2.5}
We now presents the results that our models yields when applied to daily PM$_{2.5}$ species data collected in California during year 2021. We apply our model to the log concentrations of the 6 pollutants, for which we have provided details in \Cref{sec:data}.

\begin{figure}[H] 
    \centering
    \includegraphics[width=\linewidth]{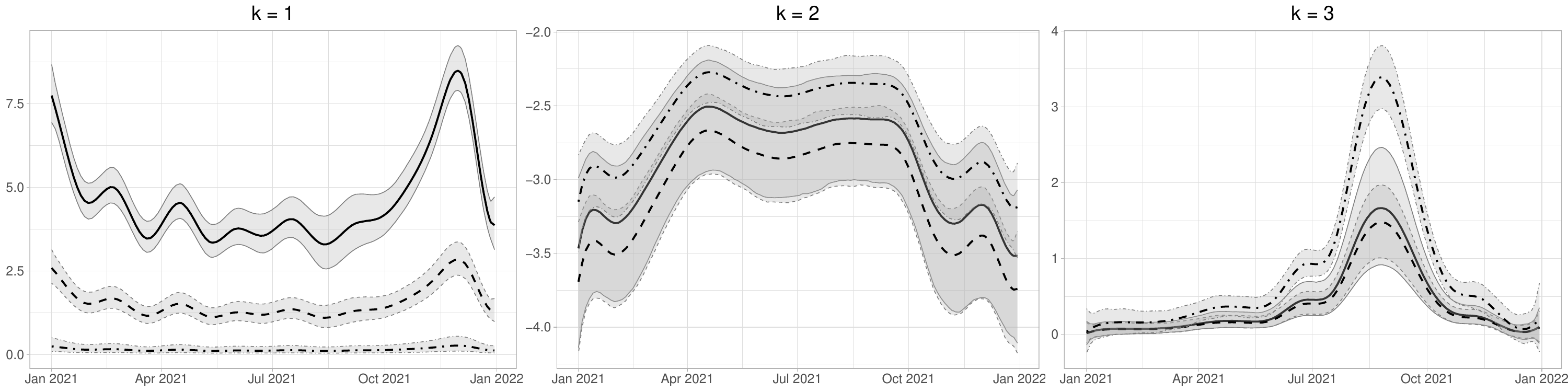}
    \caption{Posterior median and 95\% pointwise credible intervals for the local source emission profiles $g_k(\bm{s}_i,t)$, $k=1,2,3$, at three different sites: Los Angeles (solid lines), White Mountain (dot-dash lines) and Pinnacles National Park (dashed lines).}
    \label{fig:real_postG}
\end{figure}

\begin{figure}[H]
    \centering
    \includegraphics[width=.95\linewidth]{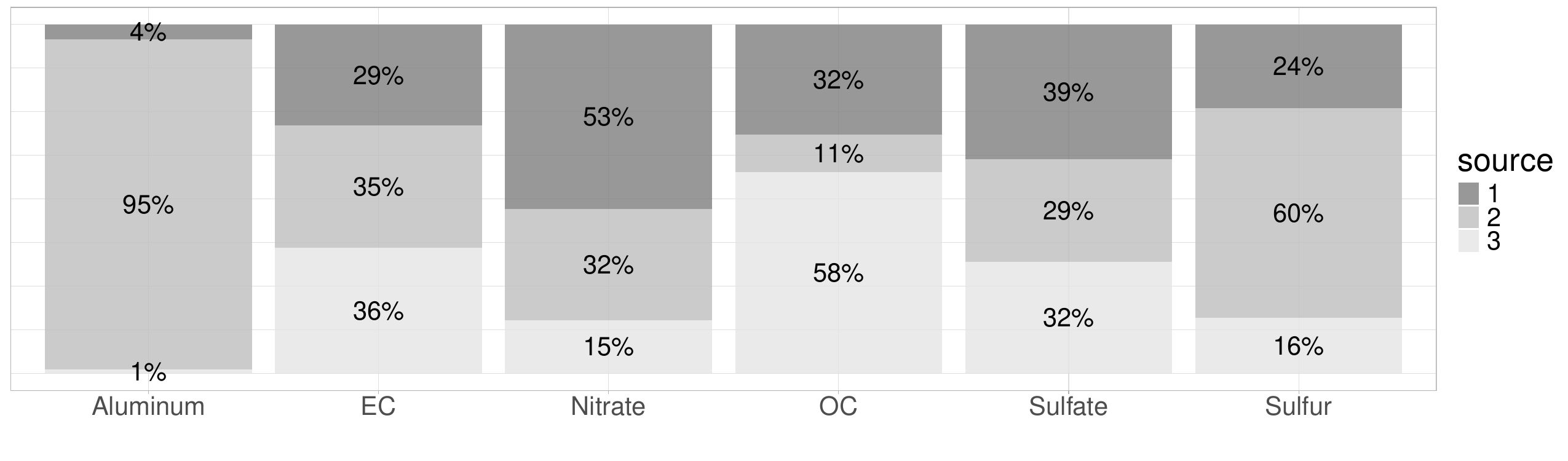}
    \caption{Estimated compositional profile $\left\{ \hat{\alpha}_{k,c}, k=1,\ldots, K^\star \right\}$ for each pollutant $c$.}
    \label{fig:real_postH}
\end{figure}

As \Cref{fig:real_postG} indicates, our model identifies $3$ sources; estimated profiles of the local source emissions, $g_k(\bm{s}_i,t)$, $t\in \mathcal{T}$, $k=1,\dots,3$, along with their point-wise 95\% credible intervals at three different locations are shown in \Cref{fig:real_postG}. The three locations are, respectively, Los Angeles (solid thick line), White Mountain (dot-dash thick line), and Pinnacles National Park (dashed thick line). 
These three locations have been selected because they possess very different site characteristics $\bm{x}_i$, which we expect would result in different estimated local emission profiles for the three sources. As our data refer to log concentrations of different components of PM$_{2.5}$, the estimated local source emission profiles can assume both positive and negative values. 
As \Cref{fig:real_postG} clearly illustrates, the local source emission profiles exhibit different trends over time depending on the source. For example, while the profiles have distinct patterns with disjoint 95\% credible intervals for source 1, they show almost identical shape and range for source 2. Examining the similarities and differences in the profiles across sites along with their temporal trends, we can advance some hypotheses on what the three sources might represent.
To be able to quantify how much each source contributes to each PM$_{2.5}$ component's concentration, we consider the following transformation of the $h_{k,c}$ values that we apply to the \textit{marginal posterior medians} $\hat{h}_{k,c}$'s: 
\begin{equation}
\label{eq:std_h}
    \hat{\alpha}_{k,c} = \frac{\hat{h}_{k,c}}{\sum_{k=1}^{K^*} \hat{h}_{k,c}} \qquad k=1,\dots,K^{\star},\ c=1,\dots,C.
\end{equation}
From \eqref{eq:std_h}, it follows that the $\hat{\alpha}_{k,c}$'s relative to a specific pollutant $c$ sum up to 1 when summed over the sources, that is, $\sum_{k=1}^{K^*} \hat{\alpha}_{k,c}=1$ for each $c=1,\dots,C$. For each pollutant $c$, $\hat{\alpha}_{k,c}$ represents the estimated percentage of pollutant $c$'s concentration that is due to source $k$. In addition, for each pollutant $c$, the $\hat{\alpha}_{k,c}$'s taken altogether as $k$ varies from 1 to $K^\star$, yield a compositional profile of the pollutant's concentration. 
\Cref{fig:real_postH} displays such compositional profile for each of the six pollutants.  

Taking into account the information gleaned from the posterior inference of the local sources emissions profiles and the source contributions, we can put forward an interpretation of the three identified sources.
Since \Cref{fig:real_postH} indicates that the majority of aluminum concentration is due to source~2, we believe that source~2 represents soil dust, as aluminum is a natural component of soil, and  it is present in elevated concentrations in California soils, particularly in Southern California \citep{silva2000}. This interpretation of source~2 appears to be corroborated also by the fact that the functional profile of the local source emissions does not vary considerably in shape from site to site and the emissions appear to be of greater magnitude during the period from April through October, when winds are stronger, the soil is drier, and dust storms are more frequent.

Conversely, \Cref{fig:real_postH} shows that, while source~1 contributes to the concentrations of all pollutants, its emissions constitute a large part of the observed nitrate and sulfate concentrations, suggesting that source~1 might represent traffic and industrial activities. Indeed, \cite{kim2019,nema2012,xin2023} observe that vehicular emissions, fossil fuel combustion and biofuel combustion are associated with elevated nitrate concentrations, whereas industrial activity and burning of fossil fuel containing sulfure tend to be associated with higher sulfate concentrations. 
Examining the local emission source profiles over time, presented in \Cref{fig:real_postG}, confirm our
interpretation. At each of the 3 sites, the local emissions from source~1 peak in the winter months and tend to be more elevated in urban areas (e.g. Los Angeles), while they assume lower values during summer months and in more natural areas, such as White Mountain and Pinnacles National Park. Such trends are consistent with traffic and industrial seasonality.  
Finally, source~3 appears to contribute mainly to elemental carbon (EC) and organic carbon (OC). As these two PM$_{2.5}$ components have large concentrations during wildfire events \citep{liang2022}, we believe that source 3 represents wildfires. Although the risk of wildfires in California is high throughout the year, it is at its highest during the months of June through October when the heat and the dryness of the soil and of the vegetation create the perfect conditions for destructive wildfires with large smoke plumes. This is reflected in the time profiles of the local emissions associated with source 3 at the three locations. In Figure~\ref{fig:real_postG}, we can notice how the emissions relative to source 3 start to increase in April to achieve their maximum value in late August-September. An inspection of the number of wildfires that occurred in California in year 2021 confirmed this trend: while 13 wildfires took place between the months of January and March 2021, the number of wildfires per month grew substantially in the following months with 10 wildfire in April, 24 in May, 51 in June, and 36 in July. On the other hand, the number of wildfires between August and December 2021 were respectively 27 in August, 14 in September, 9 in October, and 1 and 0, respectively, in November and December \citep{calfire}. 

\begin{table}[ht]
\centering
\bgroup
\def\arraystretch{1.4}
\begin{tabular*}{\columnwidth}{@{\extracolsep\fill}r|ccccc@{\extracolsep\fill}}
  \toprule
 & Rural & Suburban & Urban \& City center & Altitude \\ 
 & (baseline-intercept) & (contrast with rural) & (contrast with rural) &  \\ 
  \hline
   \quad $\bm{\beta}_1$ & 0.868 (-0.297, 1.95) & 0.82 (0.18, 1.42) & 0.74 (0.15, 1.32) & -0.57 (-1.05, -0.18) \\ 
   $\bm{\beta}_2$ & 2.35(1.03, 3.81) & -0.02 (-0.17, 0.12) & -0.09 (-0.23, 0.05) & 0.03 (-0.04, 0.11) \\ 
   $\bm{\beta}_3$ & 2.60 (1.73, 3.49) & 0.49 (-0.28, 1.27) & -0.31 (-1.14, 0.51) & 0.44 (0.03, 0.87) \\ 
   \hline
\end{tabular*}
\egroup
\vspace{0.5cm}
\caption{Posterior median and 95\% CIs for the regression coefficients $\bm{\beta}_k$, $k=1,2,3$.  
} 
\label{tab:real_betas}
\end{table}
\Cref{tab:real_betas} reports the marginal posterior median and 95\% credible intervals for the vectors of regression coefficients $\bm{\beta}_k$, $k=1,2,3$, associated to the three estimated sources. 
While the emission levels due to soil dust (source 2) and wildfires (source 3) do not seem to be different between "\emph{rural}", "\emph{suburban}" and "\emph{urban and city center}" locations, the emissions corresponding to traffic and industrial activity (source 1) are different depending on land use. Specifically, emission levels tend to be greater in suburban and urban areas compared to more rural locations. Additionally, it seems that while emissions related to wildfires are larger at higher altitude, emissions corresponding to traffic and industrial activity follow the opposite trend. This is expected as heavily forested areas are the typical locations where wildfires get started in California and they tend to have higher elevations and be less industrialized than coastal areas.

Our interpretation of the three sources appears to be supported by the estimates of the range parameters $\rho_k$ for $k=1,2,3$. The marginal posterior distribution of $\rho_2$ concentrates on high values, much larger than the maximum distance between monitoring sites, i.e. 1200 km, suggesting that the spatial random effects associated with source~2 have a spatial correlation that persist for long distances and that can be explained in light of the late summer, early Fall strong persistent winds that disperse soil dust over large territories. 
On the other hand, the marginal posterior distributions of $\rho_1$ and $\rho_3$ are concentrated around smaller values (approximately 230 km and 98 km), suggesting that the associated spatial random effects are representative of more local phenomena, such as traffic and wildfires, respectively.

\begin{figure}[h]
\centering
    \includegraphics[width=.95\linewidth]{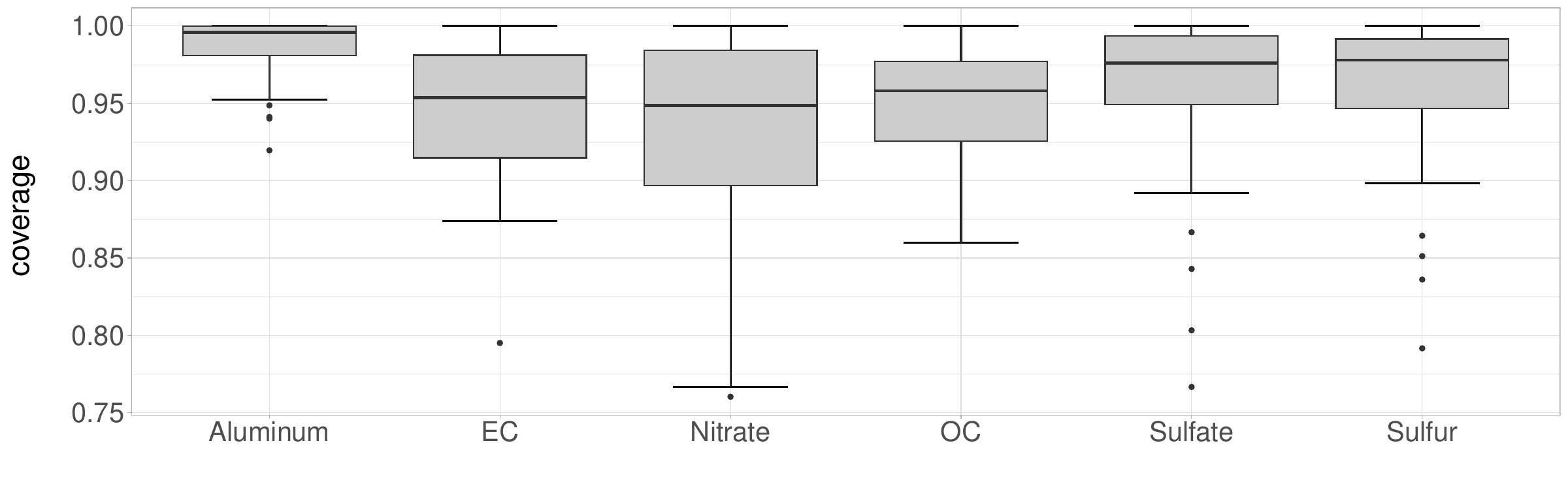}
    \caption{Boxplots showing, for each pollutant, the distribution of site-specific empirical coverages of 95\% prediction intervals of log concentrations.}
    \label{fig:real_coverage}
\end{figure}

As a final evaluation of our model, we also assess whether the posterior predictive distribution implied by our model is in concordance with the distribution of the observed data. Here for predictions we refer to in-sample predictions exclusively: due to the small number of observations locations, in fitting the model we do not hold out data from any location.
To examine the predictive performance, for each location $\bm{s}_i$, $i=1,\ldots,N$ and for each pollutant $c$, $c=1,\ldots,C$, we construct  95\% predictive intervals by simply deriving the 95\% credible intervals of the posterior predictive distribution $\tilde{y}_c(\bm{s}_i,t)$, $t\in \mathcal{T}$, $c=1,\ldots,C$. For each pollutant, we then assess whether the observed $y_{c}(\bm{s}_i,t_i)$, $i=1,\ldots,N$, $t_i\in \mathcal{T}_i$ falls in the corresponding 95\% predictive interval,  creating a binary variable denoting coverage (value equal to 1) vs. non-coverage (value equal to 0). The average of these binary variables over each set $\mathcal{T}_i$, for $1,\ldots,N$ yields a site-specific empirical coverage of the posterior predictive distribution for pollutant $c$. \Cref{fig:real_coverage} presents the distribution of the 32 site-specific empirical coverages for each pollutant $c$. As the figure shows, for all pollutants the empirical coverage is quite high and very close to the nominal level of 95\%, with some pollutants showing a slightly larger coverage than others. This is very likely a reflection of a posterior predictive distribution that is quite dispersed.

Finally, \Cref{fig:real_post_pred}
in the Appendix displays the time series of the observations of the six PM$_{2.5}$ components' concentration, on the original scale, at two distinct hold-out locations, Sacramento and Lake Tahoe Community College. These two locations have been chosen to illustrate different ranges in the observed pollution concentration levels as well as to showcase situations where our model estimates mimic the data quite faithfully versus situations where the fit is not as optimal. 

\section{Discussion}
\label{sec:discussion}
In this paper, we have offered a full Bayesian approach to solve the source apportionment problem when data on multiple air pollutants is available over a time period and from multiple locations. 
Breaking with previous work that adopted a two-stage estimation procedures, we estimate the unknown number of sources by using shrinkage priors on source-specific parameters, mimicking in part the modeling approach undertaken by \cite{montagna2012,bhattacharya2011} in other contexts. Additionally, in modeling the local emissions we have made use of information on monitoring site's characteristics, such as land use and geographical characteristics.
Application of our model to simulated data shows that we are able to retrieve the true number of sources and provide correct estimates of the target latent variables even in presence of significant amount of measurement uncertainty and variance.
Moreover, when we applied our model to PM$_{2.5}$ speciation data in California, we identified 3 sources, obtaining meaningful insights, which are consistent with California's geography and land use and which are supported by existing literature.

With respect to the current Bayesian literature on source apportionment, our model offers four novelties: $(i)$ it models the emissions from sources as spatio-temporal processes; $(ii)$ it allows for covariates to have an effect on the local emissions profiles and introduces source-specific spatial random effects to account for spatial dependence; $(iii)$ it uses a functional data approach to model the temporal dependence in the global emission profiles; and $(iv)$ it estimates the number of sources $K^*$ from the data using a multiplicative gamma process shrinkage prior. 
None of the previous Bayesian models for source apportionment accounts for spatio-temporal dependence and none estimates the number of sources using the methodologies and approaches listed above. 
We point out that adopting a functional framework to handle the temporal dependence in the data is particularly advantageous compared to an autoregressive or dynamic model formulation, that are more commonly used, \citep{jun2013multivariate,park2001,park2018}, given the irregular sampling times of PM$_{2.5}$ components. Moreover, it offers computational benefits as it bypasses the need to invert large temporal matrices in the MCMC that would be otherwise required.  
We also remark that while we elected to use the MGSP prior of \cite{bhattacharya2011} for shrinkage, alternative prior specifications can be used, such as the 
cumulative shrinkage process (CUSP) prior of \cite{legramanti2020bayesian} or its generalizations \citep{fruhwirth2023generalized}. These latter choice would allow the global emissions $f_k(t)$'s to be exactly equal to zero, which is not possible in our model.

Even though our modeling framework offers many advantages over existing methods, it also suffers from some limitations. 
For example, to render estimation of the local source emission profiles (the $g_k(\bm{s},t)$'s) feasible we have separated time and space, decomposing each $g_k(\bm{s},t)$ in a term that depends only on space and a term that depends only on time. This modeling choice constraints the local source emission profiles to have highs and lows at the same time across sites. Future work will be devoted to investigate how to model the $g_k(\bm{s},t)$'s to achieve greater flexibility while still retaining computational simplicity.

As in many examples in the literature, we do provide source apportionment for a transformation of pollutants, instead of the raw concentrations, avoiding the assumption of further constraints of the source contributions $h_{k,c}$'s \citep{park2018} which could affect the rate of convergence and the mixing of the MCMC algorithm. Using a different functional basis \citep[see][for instance]{montagna2012,pluta2024,telesca2008}
might accommodate the constraints on the source contributions, but at the cost of potential issues during the shrinkage step. We plan to investigate this aspect further.

Further work should also consider introducing meteorological covariates, which are well-known to affect the distribution of PM$_{2.5}$ over space, in the model.

\section*{Acknowledgments}
M.F. and A.G. acknowledge the support by MUR, grant Dipartimento di Eccellenza 2023-2027.

\bibliographystyle{apalike} 
\bibliography{reference}  

\newpage

\renewcommand\thesection{A\arabic{section}}
\renewcommand\thesubsection{\thesection.\arabic{subsection}}
\renewcommand\thefigure{A.\arabic{figure}} 
\renewcommand\thetable{A.\arabic{table}} 
\renewcommand\theequation{A.\arabic{equation}}
\setcounter{table}{0}
\setcounter{figure}{0}

\section*{Appendix}
This Appendix provides additional results (posterior inference on the regression coefficients) relative to the simulation study described in Section~\ref{subsec:simulation}. In addition, it also reports results for the simulated datasets generated with $\sigma^2_c=0.2, 0.3$ and $0.5$ of the empirical data variance. We also report here plots of the posterior predictive distributions of the six components of PM$_2.5$  species in California in the original scale, during the days the monitor was operational in 2021, at two distinct hold-out locations.

Table~\ref{tab:sim_beta_rho} of this Appendix reports the estimates of the regression parameters and spatial ranges in the simulation study described in 
Section~\ref{subsec:simulation}
where data where generated according to \eqref{eq:fun_bsa}
with $\sigma^2_c=0.1$ of the empirical variance of the simulated $\sum_{k=1}^{K} g_k(\bm{s}_i,t)\cdot h_{k,c}$, $i=1,\ldots,N$, $t\in\mathcal{T}$, for each $c=1,\ldots, C$.

\begin{figure}[t!]
    \centering
    \vspace*{-0.5cm}
    \subfigure[$\sigma_c^2$ set to 10\% of $y_c(\bm{s}_i,t)$'s averages.]{
    \includegraphics[width=0.85\linewidth]{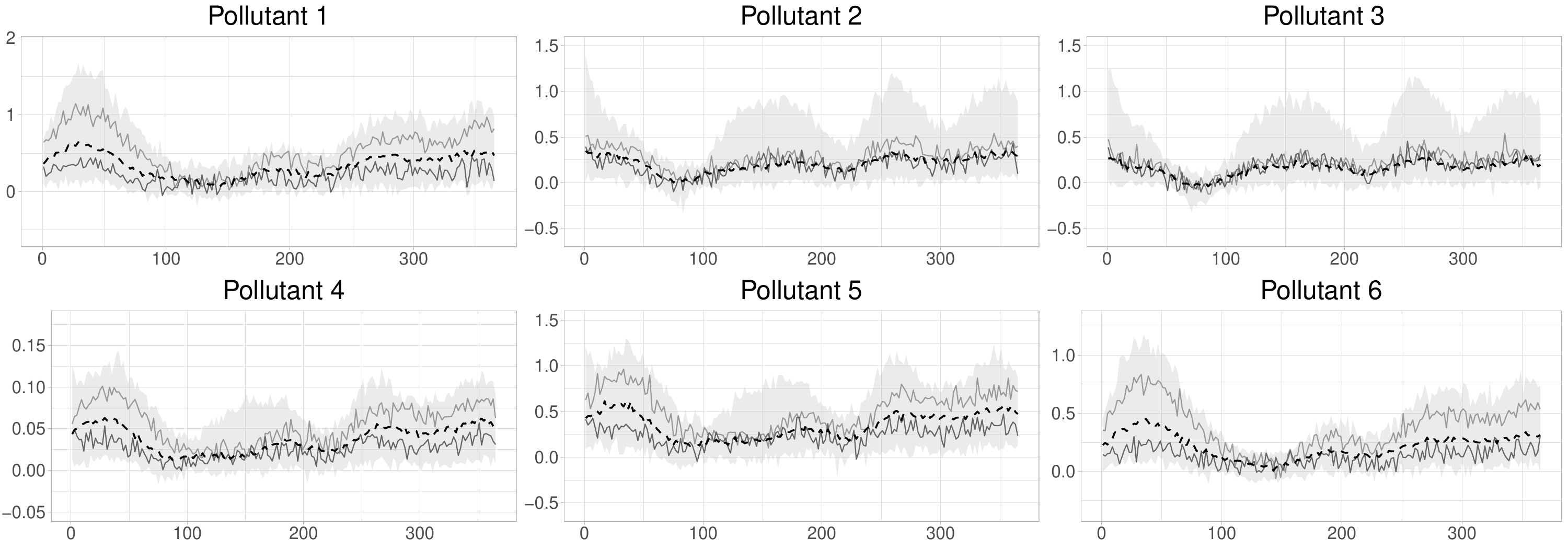} }\\
    \vspace*{-0.15cm}
    \subfigure[$\sigma_c^2$ set to 20\% of $y_c(\bm{s}_i,t)$'s averages.]{
    \includegraphics[width=0.85\linewidth]{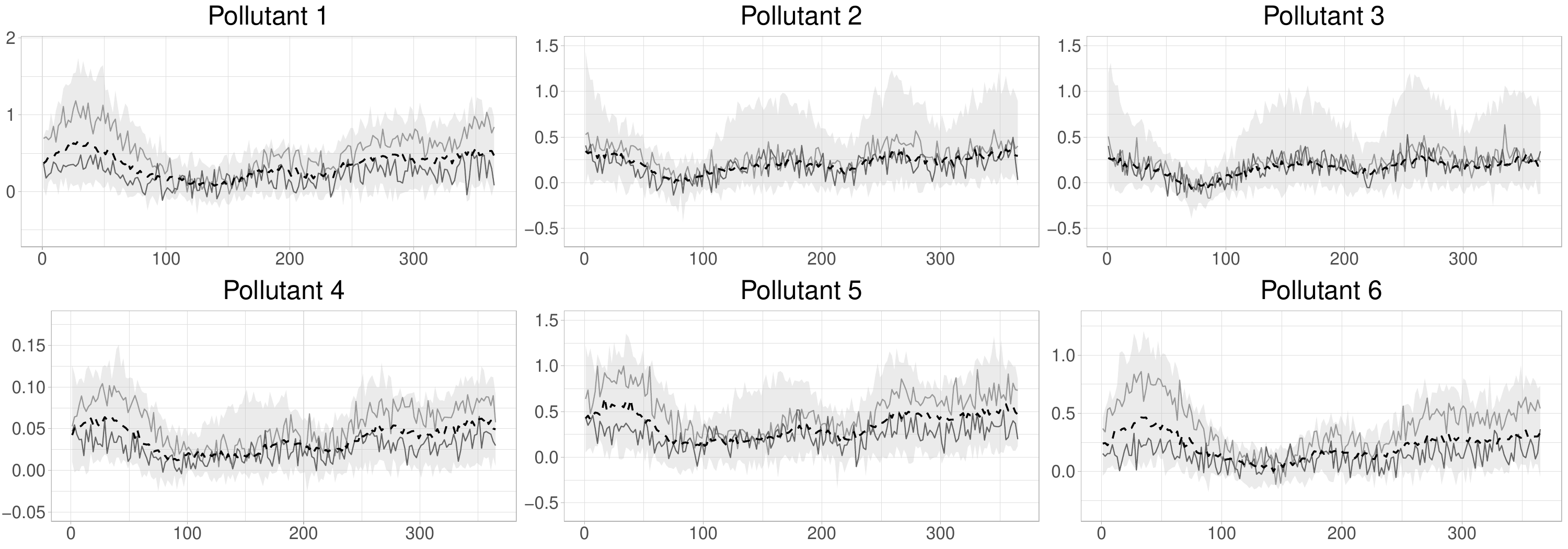} }\\
    \vspace*{-0.15cm}
    \subfigure[$\sigma_c^2$ set to 30\% of $y_c(\bm{s}_i,t)$'s averages.]{
    \includegraphics[width=0.85\linewidth]{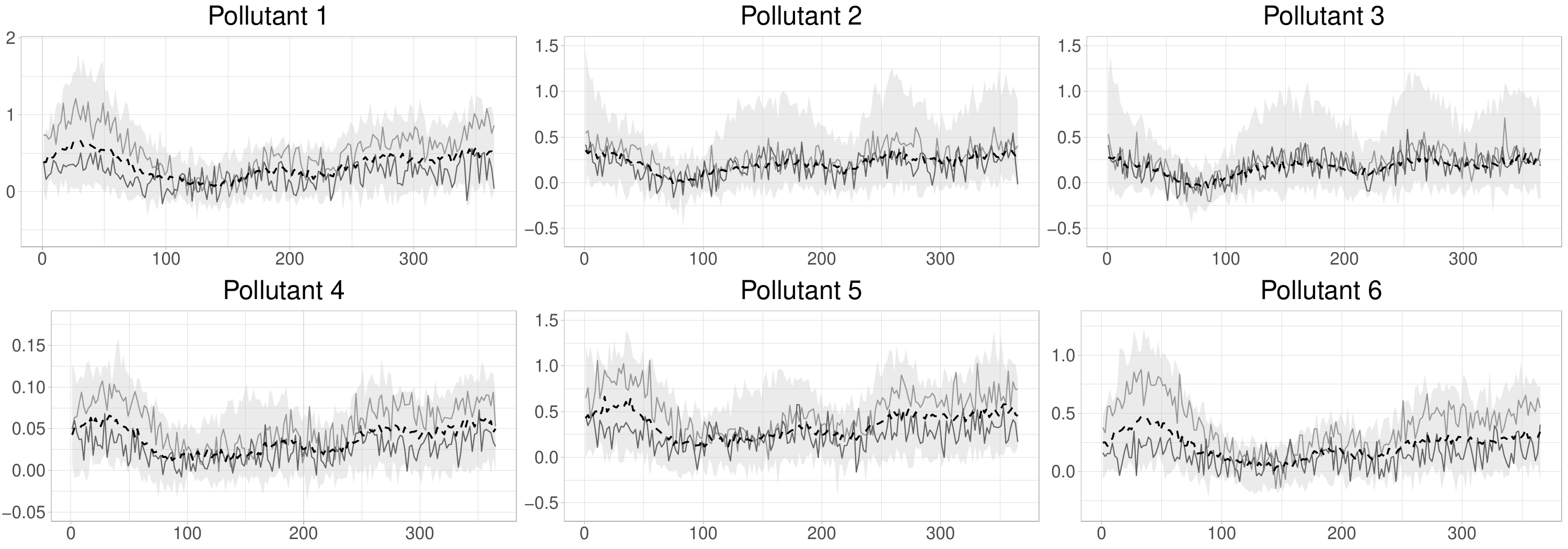} }\\
    \vspace*{-0.15cm}
    \subfigure[$\sigma_c^2$ set to 50\% of $y_c(\bm{s}_i,t)$'s averages.]{
    \includegraphics[width=0.85\linewidth]{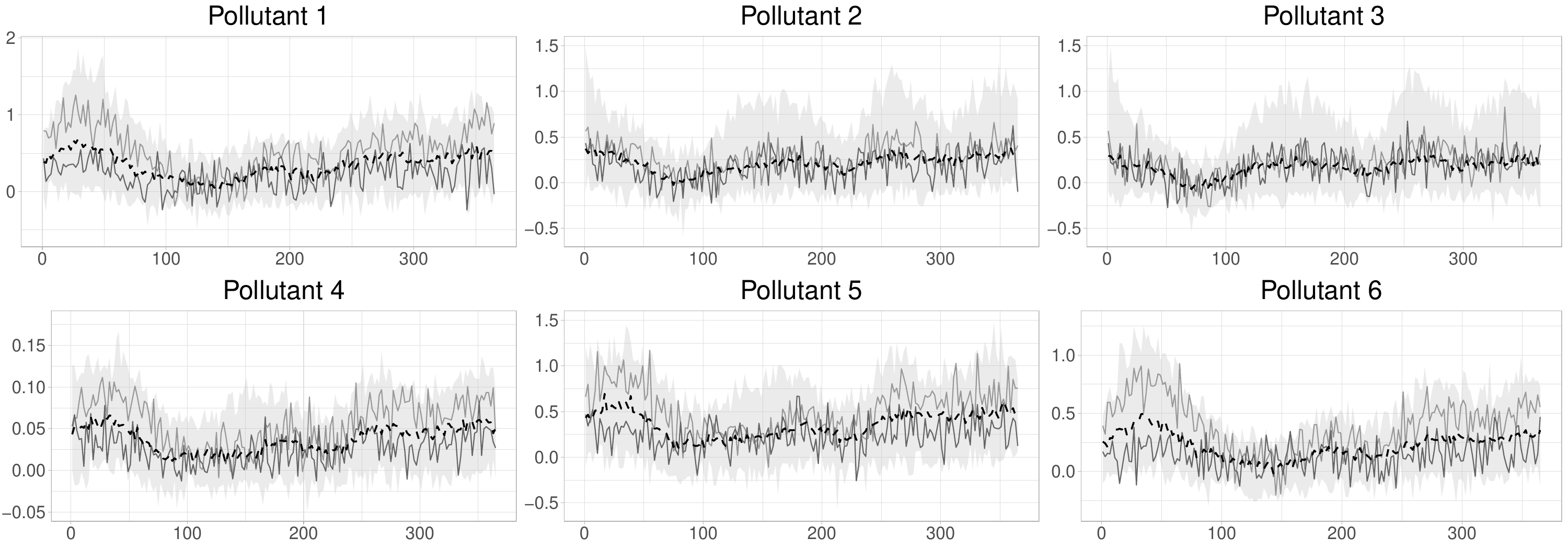} }
    \vspace*{-0.15cm}
    \caption{Simulated data $y_c(\bm{s}_i,t)$ with increasing noise level given by $\sigma_c^2$. Shaded areas represent the range of values over all sites for each time $t$. Dashed black lines represent the median value over all sites for each time $t$. Solid grey lines display the simulated data for two selected monitoring sites.}
    
   \label{fig:sim_y_var}
\end{figure}

Figure~\ref{fig:sim_y_var} shows simulated data $y_c(\bm{s}_i,t)$ from 
\eqref{eq:fun_bsa} as in Section~\ref{subsec:simulation},
but with increasing noise level given by $\sigma_c^2$ set equal to, respectively, 0,1, 0.2, 0.3 and 0.5 times the empirical variance of the simulated $\sum_{k=1}^{K} g_k(\bm{s}_i,t)\cdot h_{k,c}$, $i=1,\ldots,N$, $t\in\mathcal{T}$, for each $c=1,\ldots, C$. Note that the shaded areas represent the range of values across all sites for each $t\in \mathcal{T}$. Dashed black lines represent the median value for all sites at each time $t$. Solid grey lines display the simulated data for two selected monitoring sites. It is clear how greater values of the noise affect simulated data. 

\begin{figure}[!h]
    \centering
    \subfigure[$\sigma_c^2$ set to 20\% of $y_c(\bm{s}_i,t)$'s averages.]{
    \includegraphics[width=0.9\linewidth]{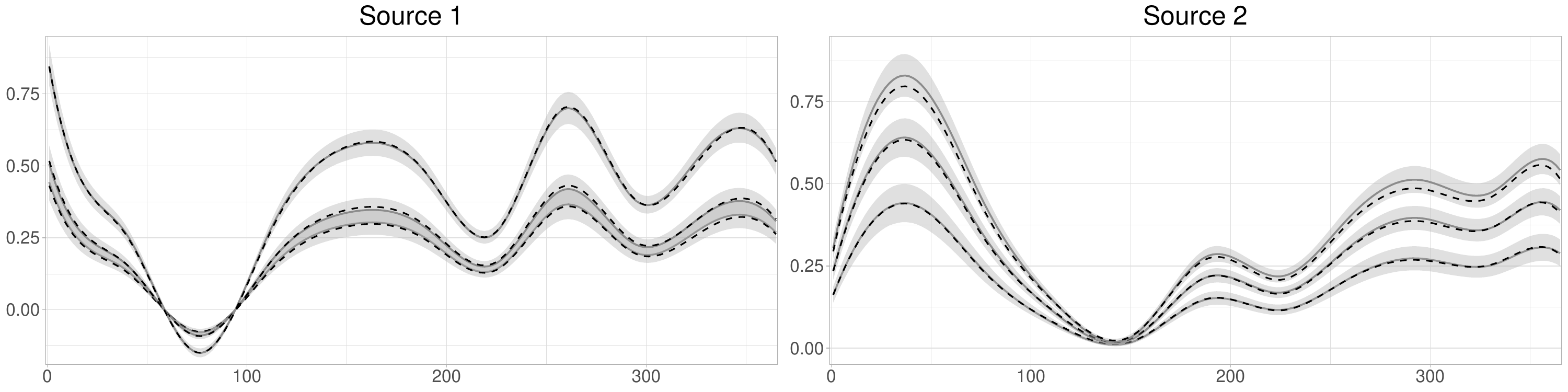} }\\
    \subfigure[$\sigma_c^2$ set to 30\% of $y_c(\bm{s}_i,t)$'s averages.]{
    \includegraphics[width=0.9\linewidth]{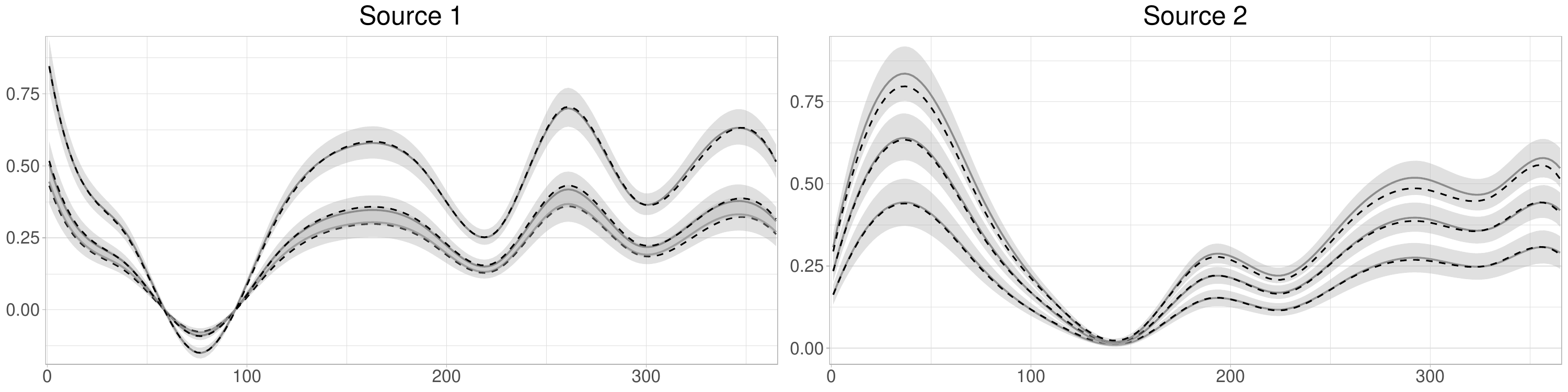} }\\
    \subfigure[$\sigma_c^2$ set to 50\% of $y_c(\bm{s}_i,t)$'s averages.]{
    \includegraphics[width=0.9\linewidth]{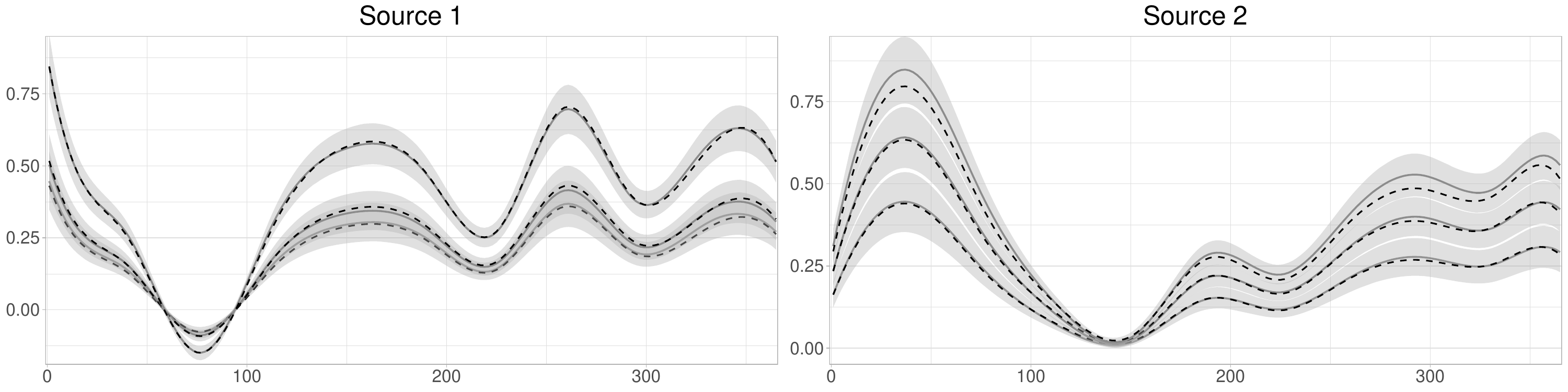} }
    \caption{Posterior median (solid line) and 95\% pointwise credible intervals (grey shaded areas) for $g_k(\bm{s}_i,t)$, $t\in \mathcal{T}$, $k=1,2$, at three representative sites $i=9,11,18$ among the $N=32$ under study. In each panel, the black dashed lines represent the true (simulated) local emission profiles. Estimates provided for different levels of noise, increasing the error variance $\sigma_c^2$ in (a) to (c).}
    
    \label{fig:sim_g_var}
\end{figure}

\Cref{fig:sim_g_var}
displays estimates of the local emission profiles $g_k(\bm{s}_i,t)$'s  for the simulated data reported in \Cref{fig:sim_y_var} with $\sigma_c^2$ set equal to, respectively, 0.2, 0.3 and 0.5 times the empirical variance of the averages of the simulated data, analogously as in Figure~\ref{fig:sim_postG}. 
\Cref{fig:sim_g_var} shows the estimated local emission profiles at three selected locations along with pointwise 95\% credible bands. For each location $\bm{s}_i$, the estimate of the local emission profile is taken to be the curve obtained by joining the posterior median of $g_k(\bm{s}_i,t)$ at each $t\in \mathcal{T}$.  Note that the model is able to recover the true number of sources $K^*=2$ for all values of  $\sigma_c^2$. 

\begin{figure}[h]
    \centering
    \subfigure[$\sigma_c^2$ set to 20\% of $y_c(\bm{s}_i,t)$'s averages.]{
    \includegraphics[width=0.85\linewidth]{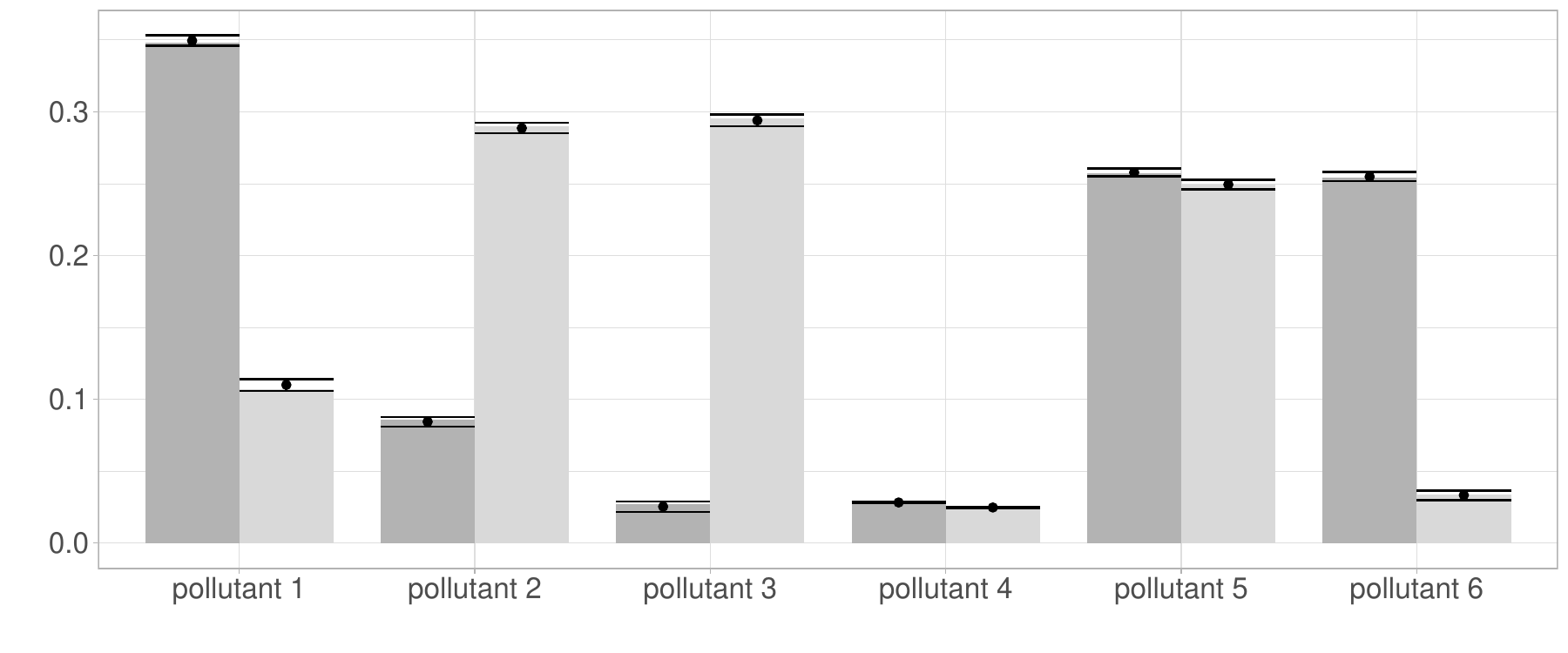} }\\
    \subfigure[$\sigma_c^2$ set to 30\% of $y_c(\bm{s}_i,t)$'s averages.]{
    \includegraphics[width=0.85\linewidth]{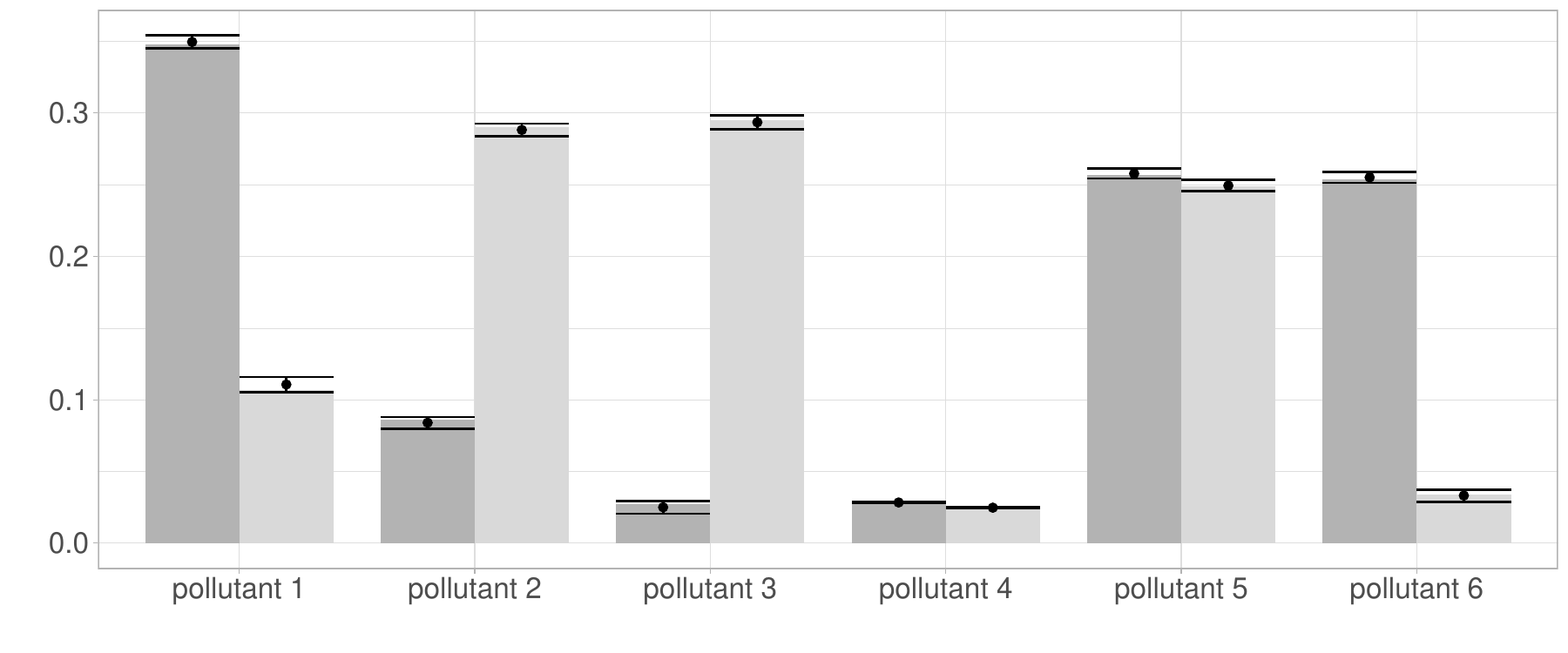} }\\
    \subfigure[$\sigma_c^2$ set to 50\% of $y_c(\bm{s}_i,t)$'s averages.]{
    \includegraphics[width=0.85\linewidth]{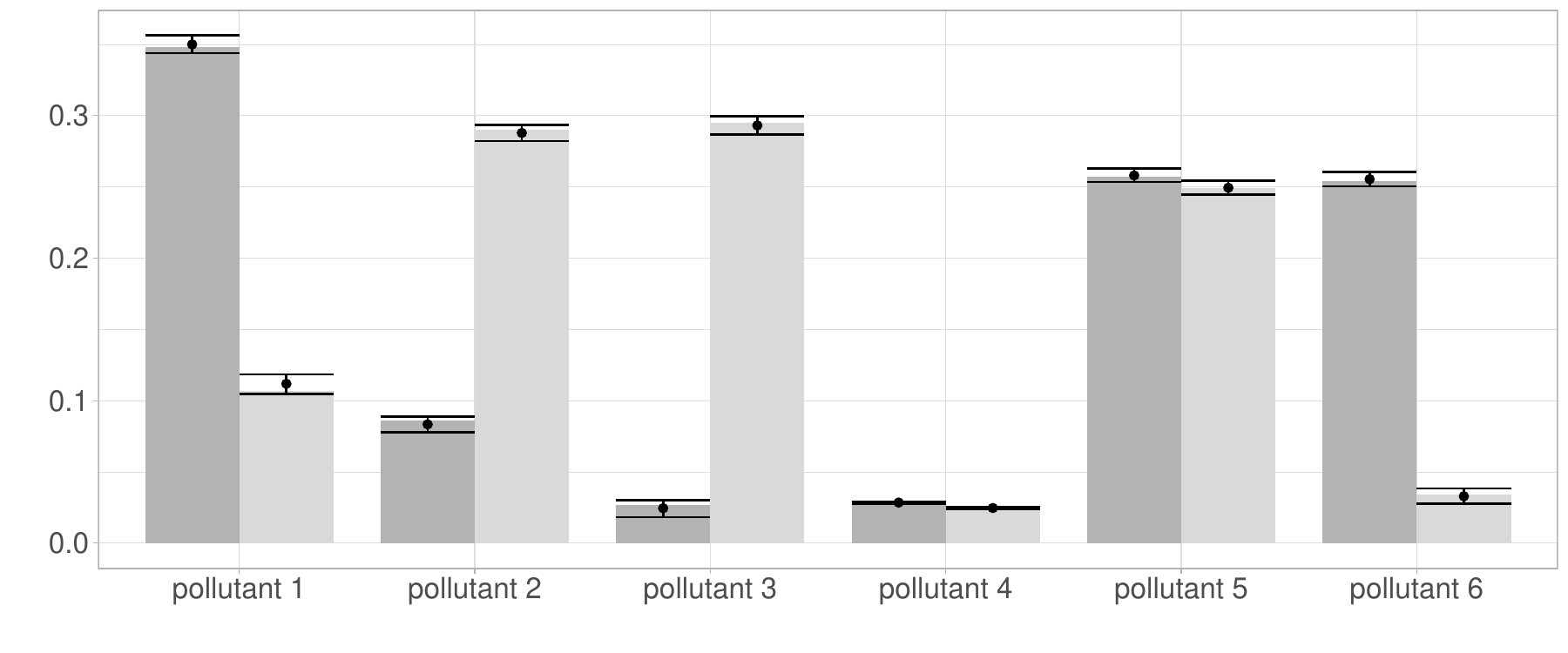} }
    \caption{Posterior median (black dot) and 95\% credible interval (error bars) of $h_{k,c}$ for $k=1,2$ and $c=1,\dots,6$. The height of the bars represent the true values of the $h_{k,c}$'s: the darker grey bars refer to the contribution of source 1 to the six pollutants concentration, while the lighter grey bars refer to source 2. Estimates provided for different levels of noise, increasing the error variance $\sigma_c^2$ in (a) to (c).}
    
    \label{fig:sim_h_var}
\end{figure}

\Cref{fig:sim_h_var} 
shows the point estimates of the $h_{k,c}$'s for each pollutant $c$ and each source $k$, i.e., the posterior medians with 95\% credible intervals for $\sigma_c^2$ set equal to, respectively, 0.2, 0.3 and 0.5 times the empirical variance of the averages of the simulated data, analogously as in \Cref{fig:sim_postH}. 

\begin{figure}[ht]
\centering
\subfigure[\emph{Sacramento -- Del Paso Manor}]{\includegraphics[width=\linewidth]{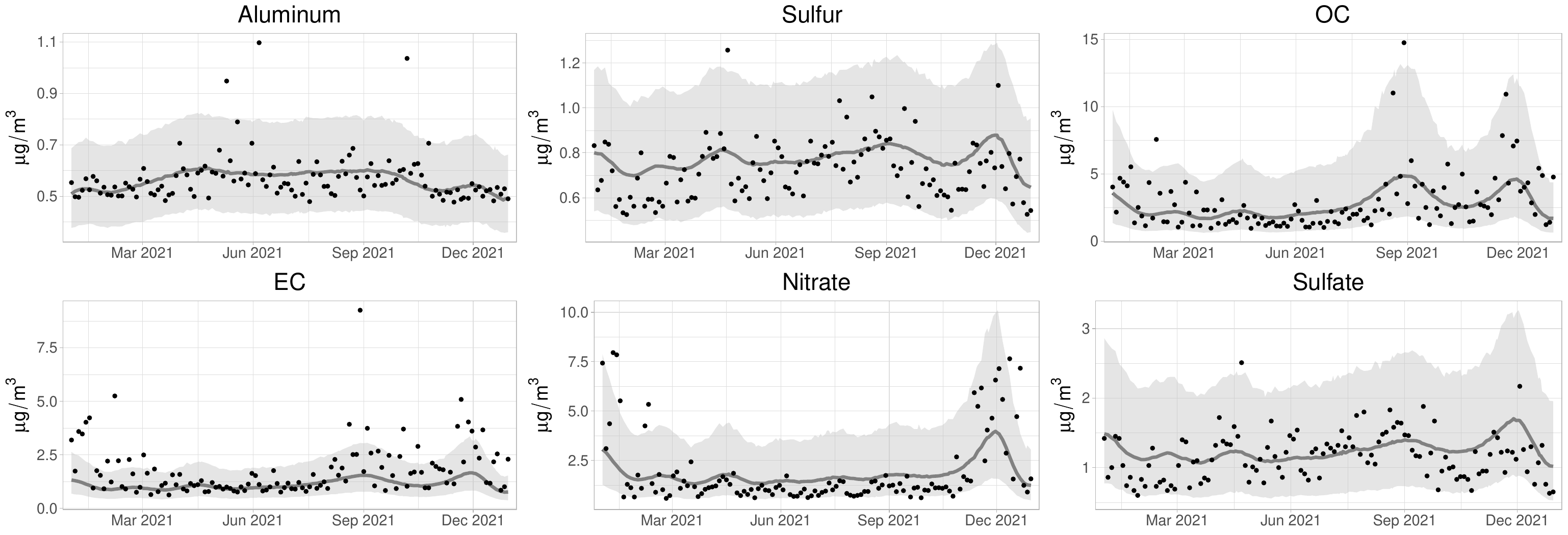}}\\
\subfigure[\emph{Lake Tahoe Community College}]{\includegraphics[width=\linewidth]{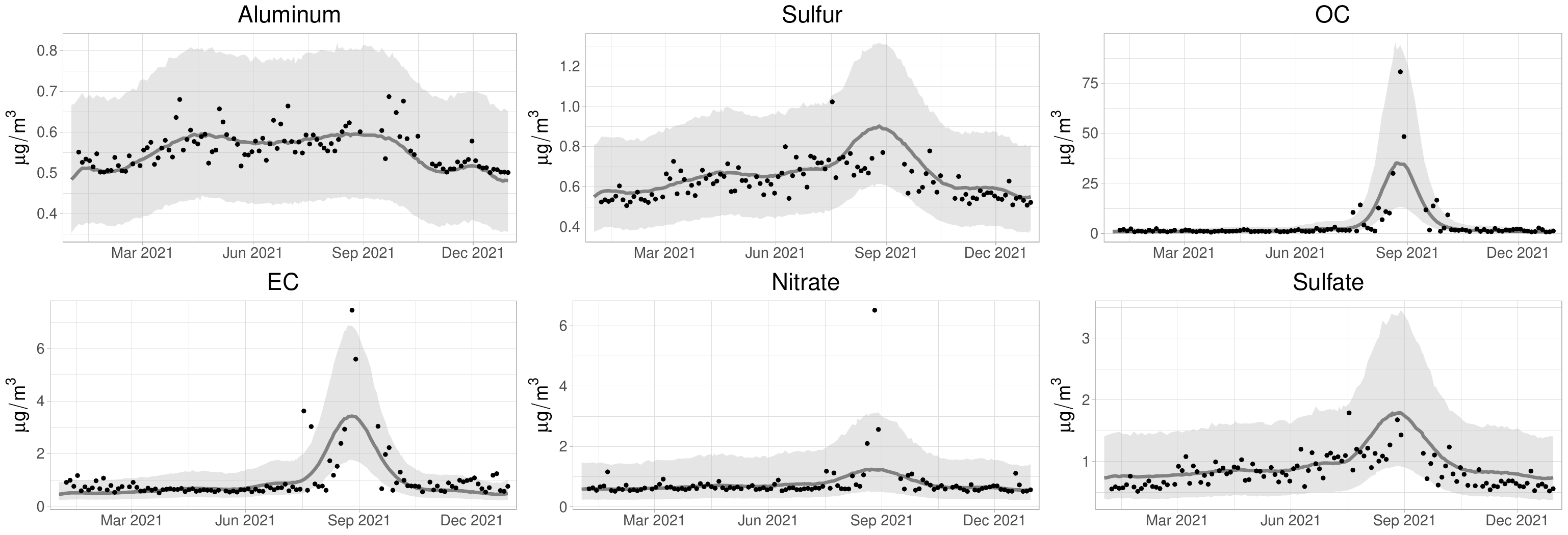}}
\caption{Observed (black points) and estimated (black solid line) daily average concentrations of the six PM$_{2.5}$ components at the two locations of \emph{Sacramento} (a) and \emph{Lake Tahoe} (b) during year 2021, presented on the original scale of $\mu g/m^3$. Observed concentrations refer only to the days the monitor at the corresponding sites were operational, whereas estimated concentrations are available for each day in 2021. 
The gray bands indicate the pointwise 95\% posterior predictive interval for each pollutant's concentration.}

\label{fig:real_post_pred}
\end{figure}

\Cref{fig:real_post_pred} displays the time series of the observations of the six PM$_{2.5}$ components' concentration, on the original scale, at two distinct hold-out locations, Sacramento and Lake Tahoe Community College. These two locations have been chosen to illustrate different ranges in the observed pollution concentration levels as well as to showcase situations where our model estimates mimic the data quite faithfully versus situations where the fit is not as optimal. 
Besides presenting, for each pollutant, the observed concentration values on the original scale, during the days the monitor was operational in 2021, the figure also presents our model's estimated concentration level for each day in the year 2021 along with the 95\% credible interval. The posterior predictive distributions are summarized using the median of the distribution. At both locations, the predicted concentration levels appear to vary in time with a certain level of smoothness, capturing the temporal dynamics of the observed concentration values.

\begin{table}[hbt!]
\centering
\begin{tabular*}{\columnwidth}
{@{\extracolsep\fill}c|lcc@{\extracolsep\fill}}
  \hline
parameter & & source $k=1$ & source $k=2$ \\
  \hline
    & rural (intercept) & 1.42 (0.33, 2.57) & 0.837 (-0.09, 1.77) \\ 
  $\bm{\beta}_k$ & suburban & -0.1 (-0.61, 0.41) & 0.15 (-0.2, 0.49) \\ 
    & urban & 0.09 (-0.40, 0.59) & 0.12 (-0.21, 0.44) \\
    & elevation & -0.03 (-0.23, 0.16) & -0.22 (-0.35, -0.1) \\
  \hline
  $\rho_k$ & & 616.89 (327.19, 1056.01) & 218.63 (124.29, 365.23) \\
  \hline
\end{tabular*}
\vspace*{0.2cm}
\caption{Marginal posterior median (and 95\% credible intervals) of the regression coefficients $\bm{\beta}_k$ and spatial parameters $\rho_k$ for the two simulated sources $k=1,2$. }
\label{tab:sim_beta_rho}
\end{table}

\newpage

\end{document}